\documentclass[12pt]{iopart}

\usepackage{cite}
\usepackage{hyperref}
\usepackage{graphicx}
\hypersetup{colorlinks=true, linktoc=all, linkcolor=blue}
\usepackage{titlesec}
\usepackage{iopams}

\expandafter\let\csname equation*\endcsname\relax
\expandafter\let\csname endequation*\endcsname\relax
\usepackage{amsmath}

\usepackage[perpage]{footmisc}

\renewcommand\contentsname{\vspace*{-40pt}}

\newcommand{\dpp}{\left(\nabla \phi \right)^2}
\newcommand{\cbr}[1]{\left(#1\right)}

\newcommand{\boxp}{\Box \phi}
\usepackage{cancel}

\newcommand{\GB}{\mathcal{G}}

\usepackage[title,toc,titletoc,page]{appendix}

\begin{document}

\title[]{A new approach and code for spinning black holes in modified gravity}

\author{Pedro G. S. Fernandes$^{1ab}$, David J. Mulryne$^{2b}$}

\address{$^a$School of Physics and Astronomy, University of Nottingham, University Park, Nottingham, NG7 2RD, United Kingdom}
\address{$^b$School of Physics and Astronomy, Queen Mary University of London, Mile End Road, London, E1 4NS, United Kingdom}
\ead{$^1$pedro.fernandes@nottingham.ac.uk, $^2$d.mulryne@qmul.ac.uk}
\vspace{10pt}

\begin{abstract}
We discuss and implement a spectral method approach to computing stationary and axisymmetric black hole solutions and their properties in modified theories of gravity. The resulting code is written in the \emph{Julia} language and is transparent and easily adapted to new settings. We test the code on both general relativity and on Einstein-Scalar-Gauss-Bonnet gravity. It is accurate and fast, converging on a spinning solution in these theories with tiny errors ($\sim \mathcal{O}\left(10^{-13}\right)$ in most cases) in a matter of seconds. 
\end{abstract}

%
%
%
%
%

\maketitle
\section*{Table of Contents}
\tableofcontents 
\newpage

\section{Introduction}
\label{sec:intro}
In the last decade, with the observation of gravitational wave events by the LIGO Scientific Collaboration \cite{LIGOScientific:2016aoc,LIGOScientific:2016sjg,LIGOScientific:2017bnn,LIGOScientific:2017vox,LIGOScientific:2017ycc,LIGOScientific:2017vwq}, and  interferometry measurements of the centre of M87 and the Milky Way by the Event Horizon Telescope Collaboration \cite{EventHorizonTelescope:2019dse,EventHorizonTelescope:2022xnr,EventHorizonTelescope:2022xqj}, we have entered a new era of testing gravity, probing the nature of black holes and Einstein's theory of general relativity (GR) in the previously inaccessible strong field regime.

\par In GR, mathematical theorems guarantee that in (electro-)vacuum the gravitational field of stationary black holes is described \textit{uniquely} by the Kerr(-Newman) metric \cite{Chrusciel:2012jk}. As eloquently put by Subrahmanijan Chandrasekhar, the uniqueness 
theorems along with a set of other results dubbed \textit{no-hair theorems} (see \cite{Herdeiro:2015waa} for a review) assert that 
the Kerr metric provides  ``the absolute exact representation of untold numbers of massive black holes that populate the universe''. While all strong regime observations are so far compatible with this ``Kerr hypothesis", any eventual deviation would provide a much sought after smoking-gun for new physics.


\par Indeed, once we go beyond GR and delve onto the realm of modified theories of gravity, 
stationary vacuum spacetimes need not to be described by the Kerr metric. Popular examples of black hole spacetimes defying the Kerr hypothesis include gravity coupled with new (complex) bosonic degrees of freedom \cite{Doneva:2022ewd,Herdeiro:2014goa,Herdeiro:2015gia,Herdeiro:2016tmi}, scalar-Gauss-Bonnet gravity \cite{Sotiriou:2013qea,Sotiriou:2014pfa,Doneva:2017bvd,Silva:2017uqg,Antoniou:2017acq,Cunha:2019dwb,Dima:2020yac,Herdeiro:2020wei,Berti:2020kgk,Kanti:1995vq,Kleihaus:2011tg,Cunha:2016wzk,Delgado:2020rev}, 4D-Einstein-Gauss-Bonnet gravity  \cite{Glavan:2019inb,Fernandes:2022zrq,Lu:2020iav,Kobayashi:2020wqy,Fernandes:2020nbq,Hennigar:2020lsl,Fernandes:2021dsb,Aoki:2020lig,Fernandes:2021ysi,Clifton:2020xhc}, and dynamical Chern-Simons gravity \cite{Alexander:2009tp,Jackiw:2003pm,Yagi:2012ya,Cano:2021rey,Cano:2019ore}.

Modification of the field equations describing gravity, however, naturally  leads to an increase in their complexity  such that analytic analysis becomes intractable. With closed-form solutions not available,  one is forced to resort either to perturbation theory or numerical methods. In the strong-field regime,  perturbative approximations may not be well-justified, leaving numerical studies as the most promising way forward. In this arena, the ever-increasing precision of our observations and measurements necessitates increasingly accurate solutions.

\par 

In this paper, we will describe a numerical method and code capable of solving with high accuracy a system of non-linear elliptic partial differential equations (PDEs), such as those that appear when analyzing stationary and axially symmetric spacetimes, and implement this in a publicly available code. 
A first version of our numerical implementation is available in the GitHub repository in Ref. \cite{gitlink}. The code is written in Julia and  can be run with ease on laptop-class computers, with solutions being found in a matter of seconds. The Julia language is fast, memory efficient, and easy to manipulate, ensuring that implementing different modified gravity theories is not a difficult task.

Our code follows similar previous numerical solvers for these spacetimes, in particular the non-publicly-available FIDISOL/CADSOL solver \cite{Solver1,Solver2,Solver3} (which has been extensively used in the literature, see e.g. \cite{Herdeiro:2014goa, Herdeiro:2015gia, Herdeiro:2016tmi,Delgado:2020rev, Kleihaus:2015aje,Kleihaus:2011tg,Herdeiro:2020wei,Berti:2020kgk,Cunha:2019dwb}) and a recent publicly available solver developed in Refs.~ \cite{Sullivan:2019vyi,Sullivan:2020zpf}. We have several motivations for writing another code. First, in this work we show that pseudospectral methods \footnote{See Ref. \cite{Dias:2015nua} for a review in the context of gravitational solutions} are ideally suited to solving the type of equations at hand. In our tailor-made implementation we therefore make use of such methods\footnote{See also \textit{Kadath} \cite{Grandclement:2009ju}, which implements a spectral methods library for theoretical physics in C++.}, while both former codes utilise finite difference methods. In contrast to the first code mentioned above, our implementation is also open source, and moreover in our bench-marking we find our code to be far more accurate as detailed further below. The code of Refs. \cite{Sullivan:2019vyi,Sullivan:2020zpf} is also significantly more accurate than that of
the FIDISOL/CADSOL solver (though the documented accuracy is still less than our own when bench-marked on the Kerr solution) and is publicly available. This code is, however, written in C, and our use of Julia leads to simple code that can easily be adapted. Our code is also considerably faster. Our overall aim is a publicly available, accurate, well documented code that is transparent and easy to use code. Furthermore, our code provides a toolbox to explore several properties of the obtained black hole solutions, rather than being only a PDE solver.

This paper is organised as follows. In section \ref{sec:spectral} we introduce the reader to pseudospectral methods and the technical machinery that will be necessary to apply them in the context of black hole physics. Next, in section \ref{sec:bhs} we will describe how we can use the aforementioned methods to solve the stationary and axisymmetric field equations for gravity, discussing the boundary conditions, coordinate compactifications, and our numerical approach. We further discuss many of the properties that can be extracted from a spinning black hole solution. Finally, in section \ref{sec:numbhs} we start by validating our methods and code against the Kerr black hole, which is known in closed form, and later use our machinery to obtain stationary and axisymmetric black holes in Einstein-scalar-Gauss-Bonnet gravity for linear and exponential couplings. We also discuss 
the accuracy of our code, and further compare with results from other codes in published literature. We work with units such that $G=c=1$.

\section{Spectral Methods}
\label{sec:spectral}

The idea behind spectral methods is to approximate a smooth solution to a system of differential or integral equations by a sum over a finite number of basis functions. In this section, we review  how this works.  Given that our aim is a clear and adaptable code,  the presentation is relatively complete, and summarises that given in John P. Boyd's book on spectral methods \cite{Boyd}, to which the reader can turn for full details (see also Ref. \cite{Dias:2015nua}). 

For simplicity, we begin with the one dimensional case and illustrate how the method finds an approximation to the smooth solution  to a differential equation, $u(x)$, with the differential equation written in the form
\begin{equation}
    \mathcal{R}\left(x,u\right) = 0,
    \label{eq:residual}
\end{equation}
where $\mathcal{R}$ is called \textit{the residual} of the system. The solution, $u(x)$, can be approximated by a finite truncated series solution $u_N(x)$ such that  
\begin{equation}
    u(x) \approx u_N(x) = \sum_{n=0}^{N-1} \alpha_n \phi_n(x),
\end{equation}
where $\{\phi_n(x)\}_{n=0}^{\infty}$ is a set of \textit{global} and \textit{orthogonal} basis functions, $\{\alpha_n\}_{n=0}^{\infty}$ is the set of \textit{spectral coefficients}, and $N$ is the \textit{resolution}. In this setup, $u_N(x)$ can be said to be a numerical solution of the system \eqref{eq:residual} if spectral coefficients are found such that the residual is below a certain prescribed tolerance. The method is therefore global rather than local, with an exponential convergence with $N$ for problems with smooth solutions. Since black hole solutions are smooth, we expect exponential convergence when come to find such solutions using spectral methods.
This is in contrast to the polynomial convergence rate of most other numerical methods, such as finite element or finite difference schemes.
Furthermore, numerical solutions obtained via a spectral method provide an analytical approximation to the problem at hand (rather than a set of approximate numerical values at a discrete number of points).

\par As noted the basis functions must be orthogonal, which implies that 
\begin{equation}
    (\phi_n, \phi_m) = c_n \delta_{m n},
\end{equation}
where the brackets represent the inner product of two functions 
$f(x)$ and $g(x)$ with respect to the weight function, $\omega(x)>0$, on the interval $[a,b]$ as
\begin{equation}
  \left(f,g\right) \equiv \int_a^b f(x) g(x) \omega(x) dx.
  \label{eq:innerprod}
\end{equation}


\par The set of basis functions used should have a number of further properties:  i) they should be easy to compute (e.g. trigonometric functions or polynomials); ii) the approximations built out of the basis functions should converge rapidly to the true solution as the resolution is increased; iii) they should be complete, which means that any solution can be represented to arbitrarily high accuracy by taking the resolution to be sufficiently high. Two commonly used sets of basis functions that obey these requirements are sines and cosines, as used in a Fourier series, and a special class of polynomials dubbed Chebyshev polynomials.

\subsection{Chebyshev Polynomials}

\par
\begin{figure}[h!]
  \centering
      \includegraphics[width=0.7\textwidth]{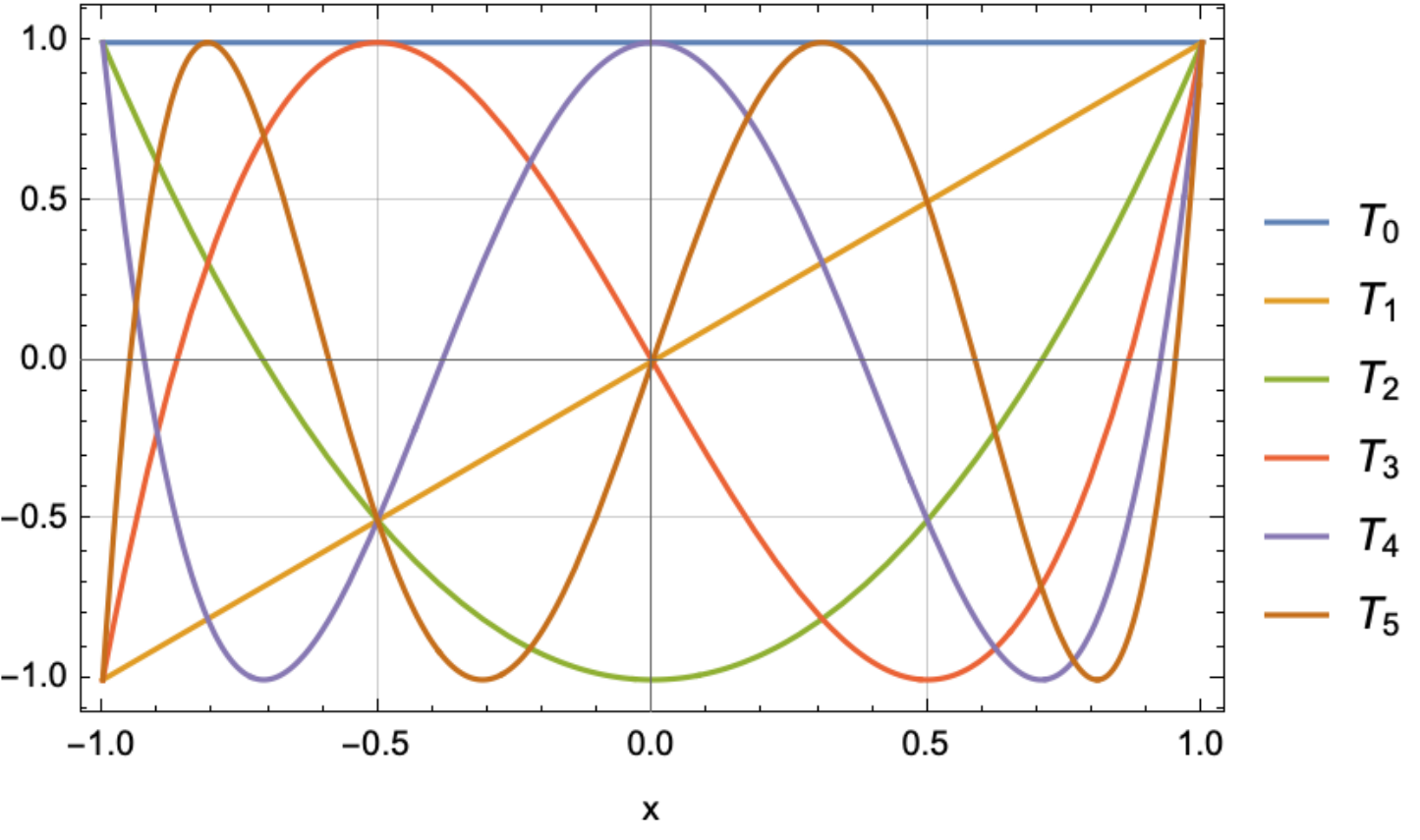}
  \caption{First six Chebyshev polynomials in the domain $x \in [-1,1]$.}
  \label{fig:chebyshevpolynomials}
\end{figure}

 For non-periodic problems, Chebyshev polynomials are the most natural choice as the spectral series is guaranteed to converge exponentially fast (provided our domain is restricted to the interval $x \in [-1,1]$). The $n$th Chebyshev polynomial (of the first kind) is defined as
\begin{equation}
    T_n\left(x\right) = \cos \left(n \theta\right), \qquad \theta = \arccos x,
\end{equation}
or equivalently by the three-term recurrence relation
\begin{equation}
    \begin{aligned}
        &T_0(x)\equiv 1, \quad T_1(x) \equiv x, \\&
        T_{n}(x) = 2xT_{n-1}(x) - T_{n-2}(x), \quad n\geq 2.
    \end{aligned}
\end{equation}
The first six Chebyshev polynomials are shown in Fig. \ref{fig:chebyshevpolynomials} in the domain $x \in [-1,1]$.

Chebyshev polynomials obey the orthogonality relation in the domain $x \in [-1,1]$
\begin{equation}
    \int_{-1}^{1} \frac{T_m(x) T_n(x)}{\sqrt{1-x^2}} dx = \frac{\pi}{2}\left(1+\delta_{0n}\right) \delta_{mn},
\end{equation}
and hence form an orthogonal basis.
Their derivatives are given by
\begin{equation}
    \frac{d}{dx}T_n(x) = n U_{n-1}(x),
\end{equation}
where $U_n(x)$ denotes the $n$th Chebyshev polynomial of the second kind, defined by the recurrence relation
\begin{equation}
    \begin{aligned}
        &U_0(x)\equiv 1, \quad U_1(x) \equiv 2x,\\&
        U_{n}(x) = 2xU_{n-1}(x) - U_{n-2}(x), \quad n\geq 2,
    \end{aligned}
\end{equation}
and with derivative
\begin{equation}
    \frac{d}{dx}U_n(x) = \frac{(n+1)T_{n+1}(x)-x U_n(x)}{x^2-1}.
\end{equation}
Note that some derivatives require special care at the boundaries $x=\pm 1$, and must be computed as a well-defined limit, namely
\begin{equation}
  \left. \frac{d^2 T_n}{dx^2} \right|_{x=-1} = \left(-1\right)^n\frac{n^4-n^2}{3}, \qquad \left. \frac{d^2 T_n}{dx^2} \right|_{x=1} = \frac{n^4-n^2}{3}.
\end{equation}

\subsection{Interpolation}

Interpolation is the process by which a function is approximated by a finite sum of suitable basis functions. The idea is that the sum is constructed such that the  
aproximation agrees with the true function at the chosen set of interpolation points (also called collocation points). The objective is that the interpolant provides a good approximation to the true function also between those points. By virtue of the minimal amplitude theorem \cite{Boyd}, Chebyshev polynomials are widely used in interpolations. The reason is twofold. First, when using the so called Chebyshev nodes (or Gauss-Chebyshev points), $x_n$, as collocation points, the effect of the Runge phenomenon (numerical instabilities near the boundaries in the form of uncontrolled oscillations) is minimized. These points are the roots of the $N$th Chebyshev polynomial, and are given by
\begin{equation}
    x_n = \cos \left( \frac{\left(2n+1\right)\pi}{2N} \right), \qquad n=0,\ldots, N-1.
    \label{eq:chebnodes}
\end{equation}
Secondly, when Chebyshev polynomials are used as the basis for the interpolation, the interpolation error is distributed uniformly over the whole range.

\begin{figure}[h!]
    \centering
    \includegraphics[width=\textwidth]{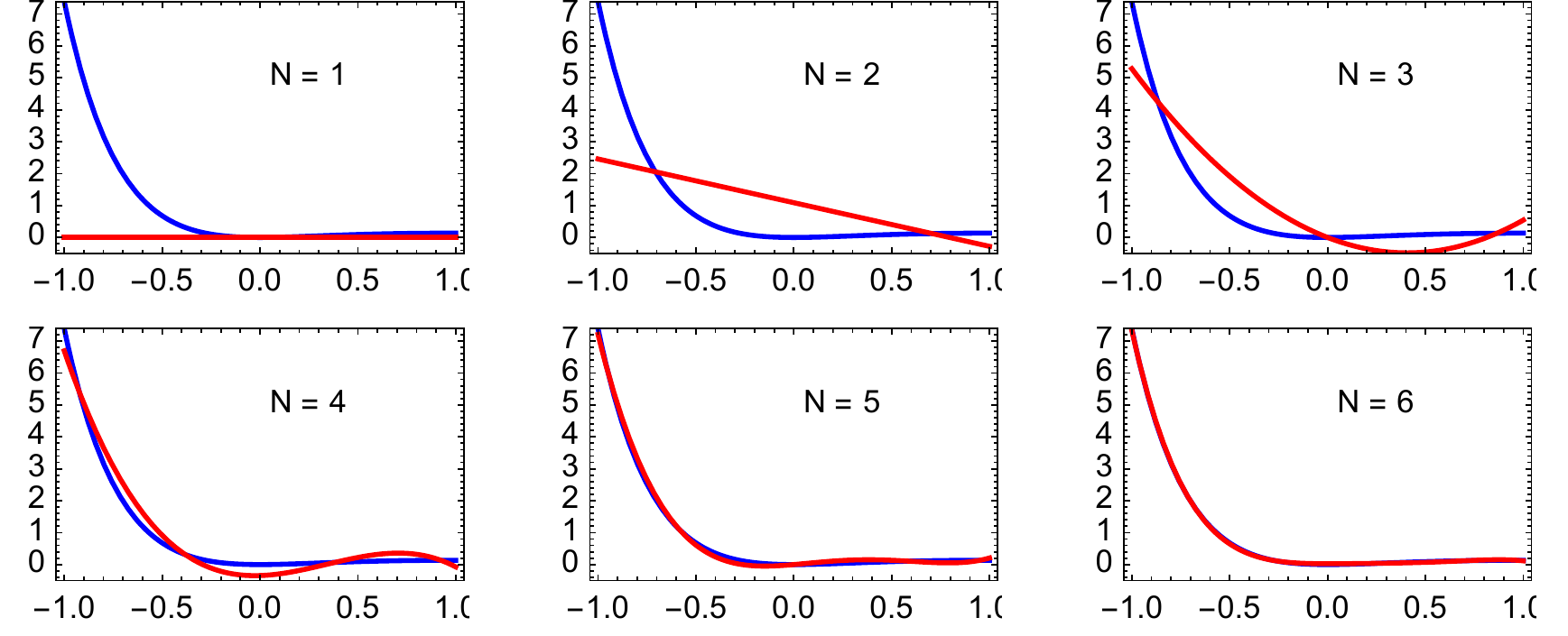}
    \caption{Interpolation of the function $u(x) = x^2 e^{-2x}$ on a Gauss-Chebyshev grid, for resolutions ranging from $N=1$ to $N=6$. Using Eqs. \eqref{eq:chebseries} and \eqref{eq:chebinterpolation} we find that for $N=6$ the spectral coefficients of the approximation $u_6(x)$ are $\alpha_0 \approx 1.48427$, $\alpha_1 \approx -2.49232$, $\alpha_2 \approx 1.85409$, $\alpha_3 \approx -1.01286$, $\alpha_4 \approx 0.395175$, and $\alpha_5 \approx -0.111169$.}
    \label{fig:interpolation}
\end{figure}

The algorithm to interpolate a smooth function $u(x)$ using a truncated Chebyshev series written as\footnote{The prime in the sum denotes that the first coefficient is halved. We chose to halve the first coefficient in the sum in order to simplify some relations below, such as Eq. \eqref{eq:chebinterpolation}.}
\begin{equation}
    u_N(x) = \frac{1}{2}\alpha_0  + \sum_{n=1}^{N-1} {\vphantom{\sum}} \alpha_n T_n(x)\equiv \sum_{n=0}^{N-1} {\vphantom{\sum}}' \alpha_n T_n(x),
    \label{eq:chebseries}
\end{equation}
relies on finding the optimal spectral coefficients $\{\alpha_n\}$, and uses the discrete orthogonality relation of Chebyshev polynomials:
\begin{equation}
    \sum_{j=0}^{N-1} T_n(x_j) T_m(x_j) = \frac{N}{2}\left(1+\delta_{0n}\right) \delta_{mn},
\end{equation}
where the $x_j$ are the points given in Eq. \eqref{eq:chebnodes}. These discrete relations imply that 
\begin{equation}
    \alpha_n = \frac{2}{N} \sum_{j=0}^{N-1} u(x_j) T_n(x_j).
    \label{eq:chebinterpolation}
\end{equation}
We present in Fig. \ref{fig:interpolation} an illustrative example of a Chebyshev interpolation performed for several resolutions using the above expressions. 

\subsection{Trigonometric functions}

For periodic problems, sines and cosines are the most suitable basis functions for a  spectral series. These obey well known orthogonality relations, and form the basis for the Fourier series representation of a periodic function. As we will see, a finite sum of these functions can be used to generate a trigonometric interpolation to a periodic function.  
Moreover, we can often simplify further by 
taking into account symmetries. For example, considering the core problem considered in this paper, we note that 
stationary and axisymmetric black holes are solutions to a system of two-dimensional elliptic PDEs that depend on the radial coordinate and the zenith angle $\theta \in [0,\pi]$. These solutions also often possess definite parity with respect to $\theta=\pi/2$ (i.e. in most cases they are symmetric about $\theta = \pi/2$), and therefore we need only to consider the range $\theta \in [0,\pi/2]$.  In this range, the following discrete orthogonality relations hold
\begin{equation}
    \begin{aligned}
        &\sum_{j=0}^{N-1} \cos\left(2n \theta_j\right) \cos\left(2m \theta_j\right) = \frac{N}{2} \left(1+\delta_{0n}\right) \delta_{mn},\\&
        \sum_{j=0}^{N-1} \cos\left(\left[2n+1\right] \theta_j\right) \cos\left(\left[2m+1\right] \theta_j\right) = \frac{N}{2} \delta_{mn},\\&
        \sum_{j=0}^{N-1} \sin\left(2n \theta_j\right) \sin\left(2m \theta_j\right) = \frac{N}{2} \left(1-\delta_{0n}\right) \delta_{mn},\\&
        \sum_{j=0}^{N-1} \sin\left(\left[2n+1\right] \theta_j\right) \sin\left(\left[2m+1\right] \theta_j\right) = \frac{N}{2} \delta_{mn},
    \end{aligned}
\end{equation}
where
\begin{equation}
\theta_n=\frac{(2n+1)\pi}{4N}, \quad n=0,\dots, N-1.
\label{eq:thetapoints}
\end{equation}
Table \ref{tab:trig} summarizes the parity properties of the functions appearing in the relations above, and 
together with these orthogonality relations  we see  that  a function, $u(\theta)$, symmetric about $\theta=0,\pi/2$ can be interpolated using only even cosines such that 
\begin{equation}
    u_N(x) =  \sum_{n=0}^{N-1} {\vphantom{\sum}}' \alpha_n \cos(2 n \theta),
\end{equation}
with the spectral coefficients
\begin{equation}
    \alpha_n = \frac{2}{N} \sum_{j=0}^{N-1} u(\theta_j) \cos(2 n \theta_j).
    \label{eq:triginterpolation}
\end{equation}

\renewcommand{\tabcolsep}{1pt}
\begin{table}[h!]
    \centering
    \begin{tabular}{|c|c|c|c|c|c|c|}
    \hline
    Fourier series & Parity w.r.t. $\theta = 0$ & Parity w.r.t. $\theta = \pi/2$ & $u(0)$ & $u(\frac{\pi}{2})$ & $\partial_\theta u(0)$ & $\partial_\theta u(\frac{\pi}{2})$ \\ \hline
    $\cos(\left[2n\right]\theta)$               & Even                       & Even                           &    $\neq 0$    &      $\neq 0$      &           $=0$             &             $=0$               \\ \hline
    $\cos(\left[2n+1\right]\theta)$               & Even                       & Odd                            &    $\neq 0$    &      $=0$      &           $= 0$             &            $\neq 0$                \\ \hline
    $\sin(\left[2n\right]\theta)$               & Odd                        & Odd                            &    $=0$    &     $=0$       &          $\neq 0$              &           $\neq 0$                 \\ \hline
    $\sin(\left[2n+1\right]\theta)$               & Odd                        & Even                           &    $=0$    &     $\neq 0$       &          $\neq 0$              &           $=0$                 \\ \hline
    \end{tabular}
    \caption{Properties of the elements of a Fourier series of a function $u(\theta)$, depending on the parity symmetries, along with a scheme of its boundary values. Here, $n \in \mathbb{N}^0$. The entries on this table for $\theta = \pi$ would be equivalent to those of $\theta=0$.}
    \label{tab:trig}
\end{table}

\subsection{Solving an ODE with a spectral method -- a first example}

\par So far we have seen how a known function can be approximated by a finite sum of suitable basis functions using interpolation. Now we turn to the problem of how to find such an approximation to an unknown function that is the solution to a given differential equation. 

To understand how to solve differential equations using a spectral method, we will first consider a simple ordinary differential equation (ODE) example. Consider the one dimensional non-linear boundary value problem
\begin{equation}
    \mathcal{R} = u_{xx}-u_{x}^2 = 0, \quad u(-1) - 2 = 0, \quad u(1) - 1 = 0.
    \label{eq:odeproblem}
\end{equation}
We will find an approximate solution to this boundary value problem in the form of a Chebyshev spectral series, and later compare our results with the known exact solution, given by
\begin{equation}
    u(x) = \log \left(\frac{2 e^2}{(e-1) x+e+1}\right).
    \label{eq:exactODE1}
\end{equation}
To illustrate the calculations analytically, we will first  consider a (very) low resolution approximate solution with $N=3$, where
\begin{equation}
        u \approx u_3 = \frac{\alpha_0}{2} T_0(x) + \alpha_1 T_1(x) + \alpha_2 T_2(x) = \frac{\alpha_0}{2} + \alpha_1 x + \alpha_2 \left(2x^2-1\right).
    \label{eq:odeexample1}
\end{equation}
Here there are $N=3$ unknowns ($\alpha_0$, $\alpha_1$, and $\alpha_2$), and $N_{BC} = 2$ boundary conditions. Once we substitute our ansatz of Eq. \eqref{eq:odeexample1} onto the residual given in Eq. \eqref{eq:odeproblem} we obtain
\begin{equation}
    \mathcal{R} \approx 4 \alpha_2-(4 \alpha_2 x+\alpha_1)^2 = 0,
    \label{eq:odeResidual}
\end{equation} 
together with the boundary conditions
\begin{equation}
    \frac{\alpha_0}{2} - \alpha_1 + \alpha_2 - 2 = 0, \qquad \frac{\alpha_0}{2} + \alpha_1 + \alpha_2 - 1 = 0.
    \label{eq:odeBC}
\end{equation}
\begin{figure}[]
  \centering
      \includegraphics[width=.85\textwidth]{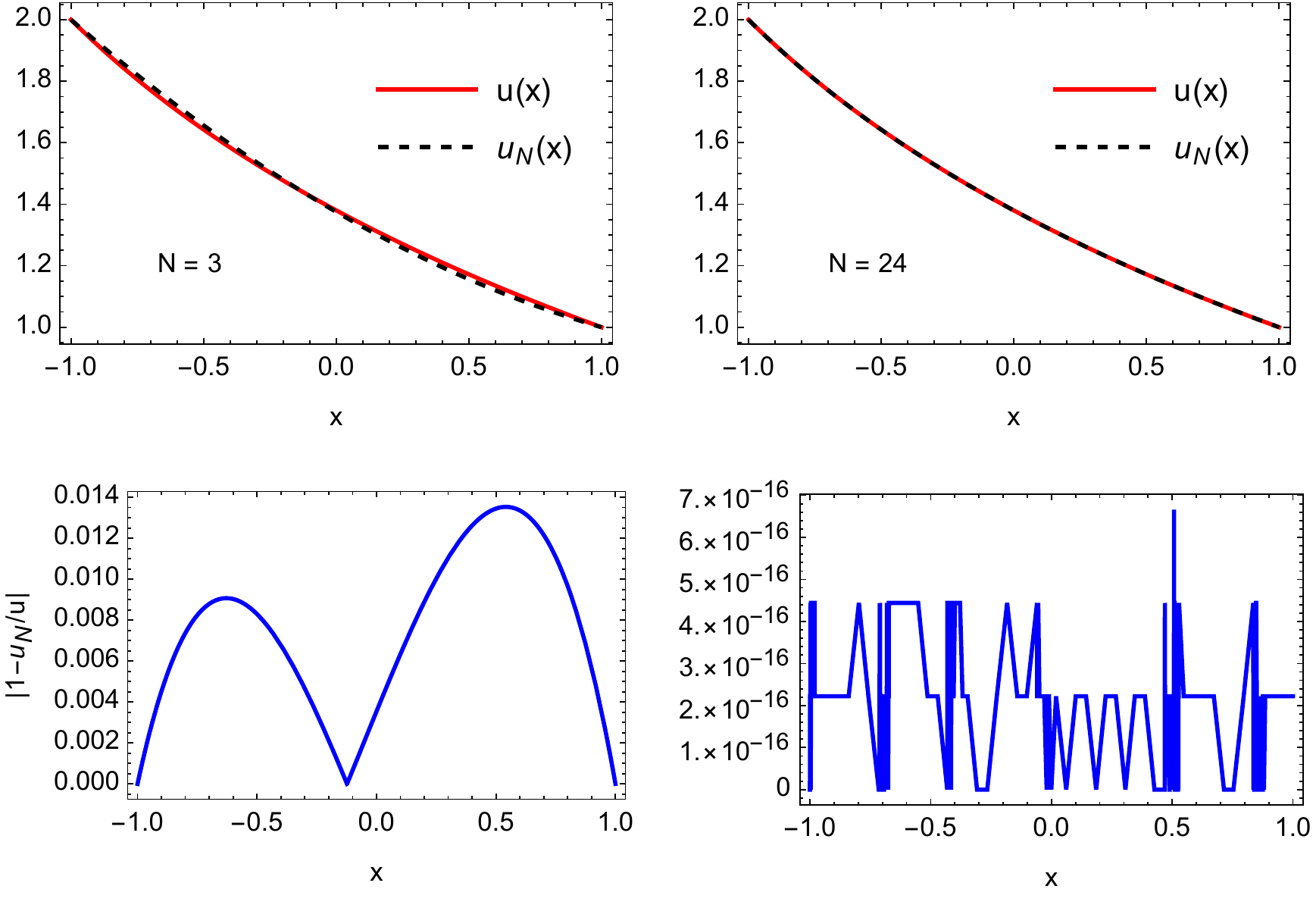}\hfill
  \caption{Approximations to the solution of the boundary value problem given in Eq. \eqref{eq:odeproblem} (top) together with their absolute errors with respect to the exact solution (bottom) for resolutions $N=3$ (left) and $N=24$ (right).}
  \label{fig:odesolution}
\end{figure} 
To find the approximate solution we simply need to determine values for the three unknowns. Given that we have only $N=3$ degrees of freedom, and the two boundary conditions provide two constraints, we need only one further equation to find the values. To get this constraint the idea is to evaluate the residual at the  $N-N_{BC}=1$ collocation point given by Eq.~\eqref{eq:chebnodes} (with the $N$ in that expression given by $1$), which gives the point $x=0$.  With our resolution of $N=3$, finding an approximate solution to the boundary value problem then reduces to solving three non-linear coupled algebraic equations for the spectral coefficients, given by the two boundary conditions of Eq. \eqref{eq:odeBC} together with the residual of Eq. \eqref{eq:odeResidual} evaluated at $x=0$. The solution to the system is
\begin{equation*}
    \alpha_0 = \frac{23}{8}, \quad \alpha_1 = -\frac{1}{2}, \quad \alpha_2 = \frac{1}{16}.
\end{equation*}
By construction this is an interpolation to the exact solution \eqref{eq:exactODE1}.

We note that had we chosen the resolution $N=4$, the number of collocation points where we would have to evaluate the residual would be $N-N_{BC}=2$, and would be given by Eq. \eqref{eq:chebnodes} as $x=\pm \sin \pi/8 \approx \pm  0.382683$. This would give  $N=4$ coupled equations to find the four unknowns in this case. This then generalises to arbitrary $N$.

\begin{figure}[]
  \centering
      \includegraphics[width=0.7\textwidth]{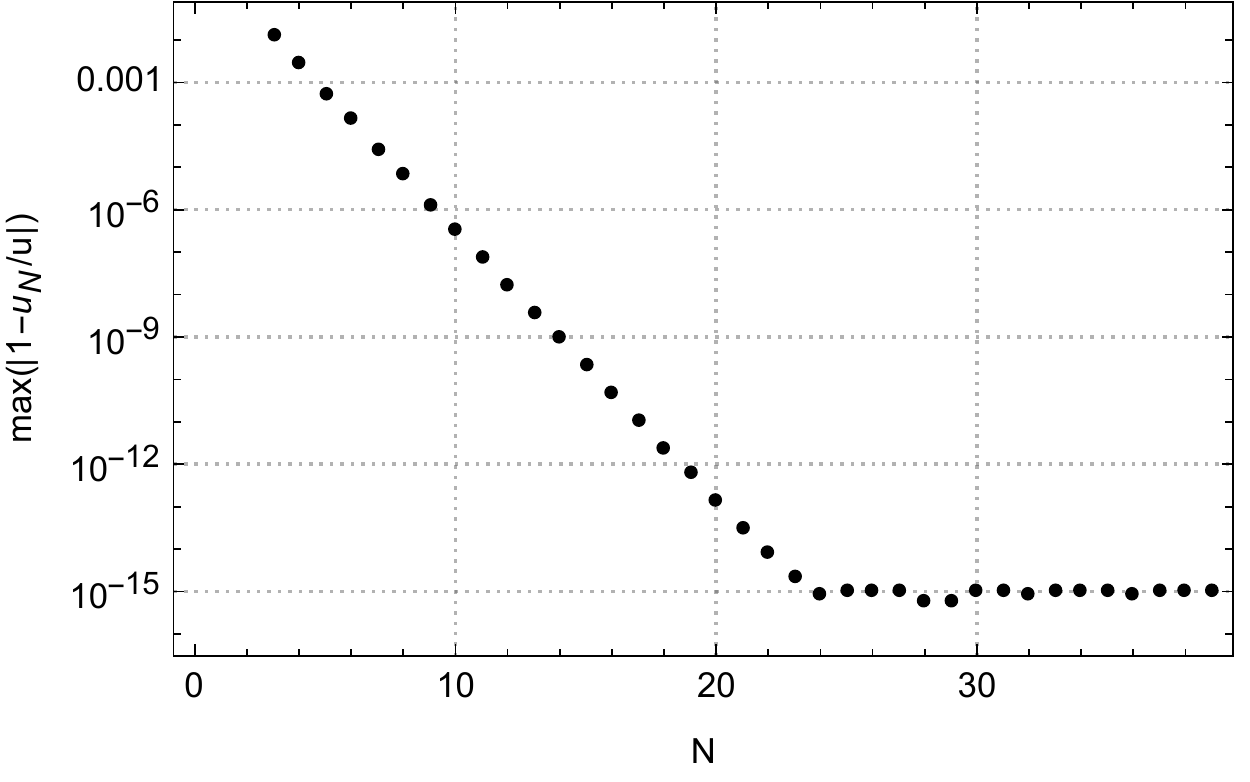}
  \caption{Logarithmic plot of the maximum absolute error in the approximation to the solution of the boundary value problem as a function of the resolution. Spectral convergence is observed, together with a roundoff plateau.}
  \label{fig:odeerror}
\end{figure}

On Fig. \ref{fig:odesolution} we plot the exact solution against the approximation obtained with $N=3$, together with the absolute error, $|1-u_N/u|$, whose maximum can be seen to be $\mathcal{O}\left(10^{-2}\right)$ already for a very low resolution $N=3$. We also plot the approximation to the solution of the boundary value problem, but for a resolution $N=24$, where we observe that errors become of order machine precision ($\mathcal{O}\left(10^{-16}\right)$). In Fig. \ref{fig:odeerror} we plot the behaviour of the maximum absolute error as a function of the resolution, where exponential convergence is observed. As a rule of thumb, the truncation error is typically the same order-of-magnitude as the last coefficient retained in the truncation series. 

\par An important point to make here is that even though for $N=3$ the approximation system has a closed-form analytical solution for the spectral coefficients, once higher resolutions are considered, a numerical root-finding method (such as Newton-Raphson) has to be employed. To successfully employ a Newton-Raphson method, a good initial guess for the spectral coefficients is of the utmost importance. We will come back to this in the next section.

\par To conclude, spectral collocation methods, also known as pseudospectral methods, are powerful tools that can be used to find high accuracy numerical solutions to differential equations. They provide global analytical approximations for the solution, and handling any kind of boundary condition is straightforward.

\subsection{Root-finding methods -- Newton-Raphson}
\par To numerically solve the system of algebraic equations for the spectral coefficients
a root-finding method must, in general, be employed. In particular, we will utilise the well  known Newton-Raphson method. 
In 
the one-dimensional case, the method attempts to solve the equation $f(x)=0$ iteratively, starting with a initial guess, $x_0$. Successive values of $x$ are then generated 
until a value, $x_*$, is reached at which the equation is approximately solved to a certain prescribed tolerance. The series of iterations takes the form
\begin{equation}
  x_{n+1} = x_n - \frac{f(x_n)}{f'(x_n)}.
  \label{eq:newton1d}
\end{equation}
For example, assume we want to find the root of the function $f(x) = x^3+x-1$, known to be $x_* \approx 0.6823278$ to eight decimal places. Starting with $x_0=1$ as our initial guess, using Eq. \eqref{eq:newton1d} we obtain
\begin{equation*}
  x_1 = 0.75, \quad x_2 = 0.68604651, \quad x_3 = 0.68233958, \quad x_4 = 0.6823278,
\end{equation*}
thus converging to $x_*$ in four iterations to the prescribed tolerance of eight decimal places. Convergence, is however, not guaranteed, and particularly in more complicated settings an appropriate choice of starting point is extremely important, and must be chosen carefully.

\par The generalization of the method to $N$ variables with $N$ equations finds the root of a vector-valued function $F:\mathbb{R}^N \to \mathbb{R}^N$, and amounts to  solving the linear system
\begin{equation}
  \mathcal{J}(\mathbf{x}_n) \left(\mathbf{x}_{n+1}-\mathbf{x}_{n}\right) = -F\left(\mathbf{x}_{n}\right),
\end{equation}
at each iteration for the unknown $\mathbf{x}_{n+1}-\mathbf{x}_{n}$, where $\mathcal{J}$ is the $N \times N$ Jacobian matrix of the system, defined as 
\begin{equation}
  \mathcal{J}_{ij} = \frac{\partial F_i}{\partial x_j}.
\end{equation}

Constructing the Jacobian matrix of a given system is not always an easy task, 
but is relatively 
straightforward for the system of equations that arises when using the spectral method to solve an ODE. Another advantage of this method. As described, in this case, the system to be solved, $F$, will be composed of the residual $\mathcal{R}$ evaluated at the Gauss-Chebyshev points \eqref{eq:chebnodes}, and the boundary conditions, and will in general involve  $u$, $u_x$ and $u_{xx}$. Our unknowns are the spectral coefficients $\alpha_j$. Thus, to facilitate the computation of the Jacobian, we may use the chain rule
\begin{equation}
  \mathcal{J}_{ij} = \frac{\partial F_i}{\partial \alpha_j} = \frac{\partial F_i}{\partial u}\frac{\partial u}{\partial \alpha_j} + \frac{\partial F_i}{\partial u_x}\frac{\partial u_x}{\partial \alpha_j} + \frac{\partial F_i}{\partial u_{xx}}\frac{\partial u_{xx}}{\partial \alpha_j},
\end{equation}
and only then substitute the spectral expansions for the function $u$. This process is easily generalizable to a system of differential ODEs/PDEs (rather than a single one)\footnote{One must be careful when labelling the spectral coefficients of the different functions as it might be a source of errors.}.

\par As previously stated, when using a Newton-Raphson method, the choice of initial guess to the spectral coefficients is extremely important, because a non-appropriate choice  will likely result in non-convergence of the algorithm. A good initial guess can sometimes be difficult to obtain, especially when dealing with systems of PDEs, where the number of coefficients is large (for our specific black holes problem, typically of $\mathcal{O}\left(10^{3}\right)$ coefficients). A good way of tackling this issue stems from a good understanding of the problem in question. For example, from an effective field theory point of view, a Kerr black hole is probably a good approximation to a black hole solution in modified theories. Therefore, since we have a closed-form expression for a Kerr black hole, an interpolation of this solution can used to generate an initial guess for the spectral coefficients in modified theories.

\section{Black Holes -- Metric ansatz, The Kerr Solution, Boundary Conditions, and Connection with the Numerical Approach}
\label{sec:bhs}
\par We will now apply the methods described in the previous sections to black hole physics and approximately solve the coupled PDEs that arise when obtaining stationary solutions in a given theory of gravity.

We will focus on a particular ansatz for the black hole spacetime written in quasi-isotropic coordinates with line-element 
\begin{equation}
  ds^2 = -f \mathcal{N}^2 dt^2 + \frac{g}{f} \left[ h \left(dr^2 + r^2 d\theta^2\right) + r^2 \sin^2\theta \left(d\varphi - \frac{W}{r}\left(1-\mathcal{N}\right) dt\right)^2\right],
  \label{eq:metric}
\end{equation}
which is stationary, axisymmetric, and circular.
Here $f$, $g$, $h$ and $W$ are dimensionless functions of the radial and angular coordinates $r$ and $\theta$, and
\begin{equation*}
  \mathcal{N} \equiv \mathcal{N}(r) = 1-\frac{r_H}{r},
\end{equation*}
where $r_H$ is the (coordinate) location of the event horizon. The spatial coordinates range over the intervals
\begin{equation}
  r \in [r_H,\infty[, \qquad \theta \in [0,\pi], \qquad \varphi \in [0,2\pi].
\end{equation}
In order for the line-element to be a solution to the theory of gravity at hand, the functions, $f$, $g$, $h$ and $W$ must satisfy a set of PDEs that result from the field equations of the theory.

The spacetime presented possesses two Killing vector fields, $k=\partial_t$ and $\Phi = \partial_\varphi$, and the linear combination
\begin{equation}
  \xi = \partial_t + \Omega_H \partial_\varphi,
  \label{eq:killingvector1}
\end{equation}
where $\Omega_H$ is the angular velocity of the horizon (to be defined below), is orthogonal to and null on the event horizon.
This Lewis-Papapetrou form for the metric is motivated by the discussion of Ref. \cite{Xie:2021bur}, which asserts that the above metric ansatz is consistent for a generic theory of gravity provided that its solutions can be obtained perturbatively from a solution in the general relativity limit. Note that our form of the metric functions on the line element of Eq. \eqref{eq:metric} differ somewhat from the standard form used in other works (see e.g. \cite{Herdeiro:2014goa, Herdeiro:2015gia, Herdeiro:2016tmi,Delgado:2020rev, Kleihaus:2015aje,Kleihaus:2011tg,Herdeiro:2020wei,Berti:2020kgk,Cunha:2019dwb}). The reasons for this will become clearer  once we make a connection to our numerical approach, and are related to numerical accuracy issues.

\subsection{General Relativity -- The Kerr Black Hole}
To begin, let us consider the known Kerr black hole, which is the solution to the stationary and axisymmetric field equations of GR in vacuum. For completeness, we present its charged generalization, the Kerr-Newman solution of electrovacuum in Appendix~\ref{ap:KN}. The Kerr black hole solves the field equations
\begin{equation}
  G_{\mu \nu} = 0,
\end{equation}
where $G_{\mu \nu}$ is the Einstein tensor, which follow from the Einstein-Hilbert action 
\begin{equation}
    \mathcal{S} = \frac{1}{16\pi} \int d^4x \sqrt{-g} R,
\end{equation}
where $R$ is the Ricci scalar of the metric $g_{\mu \nu}$. With the ansatz of Eq. \eqref{eq:metric} the Kerr black hole solution reads
\begin{equation}
    \begin{aligned}
        &f = \left(1+\frac{r_H}{r}\right)^2 \frac{\mathcal{A}}{\mathcal{B}},\\&
        g = \left(1+\frac{r_H}{r}\right)^2,\\&
        h = \frac{\mathcal{A}^2}{\mathcal{B}},\\&
        W = \frac{2M\left(M r+r^2+r_H^2\right)}{r_H r^3 \mathcal{B}}\sqrt{M^2-4 r_H^2}
    \end{aligned}
  \label{eq:kerr}
\end{equation}
where
\begin{equation}
  \begin{aligned}
    &\mathcal{A} = \frac{2 M r \left(M r+\left(r^2+r_H^2\right)\right)+\left(r^2-r_H^2\right)^2}{r^4}-\frac{\left(M^2-4 r_H^2\right)}{r^2}\sin^2\theta ,\\&
    \mathcal{B} = \left(\mathcal{A} + \frac{ \left(M^2-4 r_H^2\right) }{r^2} \sin^2\theta\right)^2 - \frac{\left(r^2-r_H^2\right)^2 \left(M^2-4 r_H^2\right)}{r^6} \sin^2\theta,
  \end{aligned}
\end{equation}
and $M$ is the ADM mass of the black hole. The total angular momentum per unit mass, $a$, of the solution is related to $M$ and $r_H$ via
\begin{equation}
  r_H = \frac{\sqrt{M^2-a^2}}{2} \equiv \frac{M}{2}\sqrt{1-\chi^2},
  \label{eq:rh}
\end{equation}
where we have defined the dimensionless spin
\begin{equation}
  \chi \equiv a/M = J/M^2.
\end{equation}
The mass $M$ and total angular momentum $J$ can be read 
off from the metric components as $r \to \infty$, where
\begin{equation}
    \begin{aligned}
        &g_{tt} = -f \mathcal{N}^2 + \frac{g \left(1-\mathcal{N}\right)^2 W^2}{f} \sin^2\theta = -1+\frac{2M}{r} + \mathcal{O}\left(r^{-2}\right), \\& 
        g_{t\varphi} = -\frac{g r \left(1-\mathcal{N}\right) W}{f}\sin^2 \theta = -\frac{2J}{r}\sin^2 \theta + \mathcal{O}\left(r^{-2}\right),
    \end{aligned}
  \label{eq:asymptoticexpansionsmet}
\end{equation}
leading to
\begin{equation}
    \begin{aligned}
        &f = 1-\frac{2\left(M-r_H\right)}{r} + \mathcal{O}\left(r^{-2}\right), \\&
        W = \frac{2J}{r_H r}+ \mathcal{O}\left(r^{-2}\right).
    \end{aligned}
  \label{eq:asymptoticexpansions}
\end{equation}
Note that the Kerr black hole in the quasi-isotropic coordinate system presented in Eq. \eqref{eq:metric} can be obtained from the standard textbook Boyer-Lindquist coordinates solution with the radial coordinate transformation
\begin{equation}
  r_{BL} = r + M + \frac{M^2 - a^2}{4r} = r \left(1 + \frac{M}{r} + \frac{r_H^2}{r^2}\right).
  \label{eq:coord_transf}
\end{equation}
The inverse transformation is given by
\begin{equation}
  r = \frac{1}{2}\left(r_{BL} - M + \sqrt{\left(r_{BL}-M\right)^2 - 4 r_H^2}\right).
  \label{eq:coord_transf_inv}
\end{equation}

\subsection{Boundary Conditions}
\par To solve the set of PDEs that result from the field equations in a particular theory of gravity, suitable boundary conditions should be imposed. These are obvious if an exact solution, such as the Kerr solution, is known by a trivial examination of the metric functions. However, in more intricate cases in modified gravity lacking an exact solution the boundary conditions must be found with a careful examination of the field equations and employing suitable expansions of the involved functions near the domain boundaries. For example if theories possess a GR limit when some parameter tends to zero, an expansion about the Kerr solution is possible. With this process, we find that in all cases to be discussed in this work within modified gravity theories, the metric functions must obey the same boundary conditions as the Kerr solution does. 
These conditions are summarized next.

\par \textbf{(i) Axis boundary conditions:} Axial symmetry and regularity of the solutions on the symmetry axis $\theta=0, \pi$, imply the following boundary conditions
\begin{equation}
  \partial_\theta f = \partial_\theta g = \partial_\theta h = \partial_\theta W = 0, \qquad \mathrm{for} \quad \theta=0, \pi.
\end{equation}
Moreover, the absence of conical singularities further imposes that on the symmetry axis
\begin{equation}
  h = 1, \qquad \mathrm{for} \quad \theta=0, \pi.
  \label{eq:conicalsing}
\end{equation}
All solutions to be discussed in this work 
are are also symmetric with respect to a reflection on the equatorial plane $\theta=\pi/2$. Therefore, as was discussed above, it is enough to consider the range $\theta \in [0,\pi/2]$ and one of the boundary conditions becomes
\begin{equation}
  \partial_\theta f = \partial_\theta g = \partial_\theta h = \partial_\theta W = 0, \qquad \mathrm{for} \quad \theta=\pi/2.
\end{equation}

\par \textbf{(ii) Event horizon boundary conditions:} The black hole solutions discussed here possess an event horizon located at a surface with constant radial variable $r=r_H$. The boundary conditions that the metric functions $f$, $g$ and $h$ obey at $r=r_H$ are
\begin{equation}
    \begin{aligned}
        &f-r_H \partial_r f = 0 \\&
        g+r_H \partial_r g = 0, \\&
        \partial_r h = 0.
    \end{aligned}
\end{equation}
The reason for the Robin-type boundary conditions that the functions $f$ and $g$ obey comes from the inclusion of the $\mathcal{N}^2$ factor in front of $f$ in the coefficient that multiplies $dt^2$ in the metric ansatz, Eq. \eqref{eq:metric}. This factor is chosen such that these functions do not contain a double-zero in a near-horizon expansion, allowing for more accurate solutions in this region, and therefore, a more accurate extraction of horizon physical quantities such as the area and temperature of the event horizon. We find that there are (at least) two possibilities for the condition that the function $W$ should obey at the horizon, one of which must be chosen appropriately such that the number of input parameters is kept at two\footnote{The ``input parameters" are the parameters needed to uniquely define a solution, this is discussed fully below in Sec. \ref{sec:connectio_numerical_approach}}
\begin{equation}
  W = r_H \Omega_H
  \label{eq:WBC_H_1}
\end{equation}
or
\begin{equation}
  W - \frac{r_H}{2} \partial_r W = 0,
\end{equation}
where $\Omega_H$ is a constant interpreted as the angular velocity of the event horizon, which in the case of a Kerr black hole is given by
\begin{equation}
  \Omega_H^{\mathrm{Kerr}} = \frac{\sqrt{M^2-4 r_H^2}}{2M \left(M+2r_H\right)} = \frac{\chi^2 - 1 + \sqrt{1-\chi^2}}{4r_H \chi}.
  \label{eq:omh}
\end{equation}

\par \textbf{(iii) Asymptotic boundary conditions:} Requiring asymptotic flatness (i.e., that as $r \to \infty$, our solution approaches the Minkowski spacetime), the functions $f$, $g$, and $h$ obey
\begin{equation}
  \lim_{r \to \infty} f = \lim_{r \to \infty} g = \lim_{r \to \infty} h = 1.
\end{equation}
Similarly to the boundary conditions at the event horizon, we find (at least) two suitable conditions for the function $W$
\begin{equation} 
  \lim_{r \to \infty} W = 0,
\end{equation}
or, from the asymptotic expansion of Eq. \eqref{eq:asymptoticexpansions}
\begin{equation}
  \lim_{r \to \infty} r_H r^2 \partial_r W + 2 M^2 \chi = 0 \Leftrightarrow \lim_{r \to \infty} \frac{r^2}{2r_H} \partial_r W + \left(1+\frac{r^2}{2r_H} \partial_r f\right)^2 \chi = 0.
\end{equation}

\subsection{Connection with the numerical approach}
\label{sec:connectio_numerical_approach}
\par To recap, the field equations of a gravitational theory once applied to the line element of Eq. \eqref{eq:metric} will result in a set of non-linear coupled elliptic PDEs in $r$ and $\theta$ subject to the boundary conditions described above. Our objective is to solve this system of PDEs numerically using a spectral method. For this we introduce the compactified radial coordinate
\begin{equation}
  x = 1-\frac{2r_H}{r},
  \label{eq:x}
\end{equation}
mapping the range $r\in [r_H, \infty[$ to
\begin{equation}
  x \in [-1,1].
\end{equation}
With the compactified coordinate, the radial boundary conditions change, and we proceed to give the new conditions next.

\noindent\textbf{Event horizon boundary conditions:} The boundary conditions that the metric functions $f$, $g$ and $h$ now obey are
\begin{equation}
    \begin{aligned}
        &f-2 \partial_x f = 0, \\&
        g+2 \partial_x g = 0,\\&
        \partial_x h = 0,
    \end{aligned}
\end{equation}
for $x=-1$.
For the function $W$, the first possibility (Eq. \eqref{eq:WBC_H_1}) remains unchanged ($W|_{x=1} = r_H \Omega_H$), whereas the second becomes
\begin{equation}
  W - \partial_x W = 0,
  \label{eq:W_BC_CHI_H}
\end{equation}
at $x=-1$.

\noindent\textbf{Asymptotic boundary conditions:} The asymptotic boundary conditions the functions $f$, $g$, and $h$ are now
\begin{equation}
  f=g=h=1, \qquad \mathrm{for} \quad x=1
\end{equation}
Asymptotically, function $W$ now obeys either
\begin{equation}
  W = 0,
  \label{eq:W0_BC_x1}
\end{equation}
or
\begin{equation}
  \partial_x W + \left(1 + \partial_x f\right)^2 \chi = 0,
  \label{eq:W_BC_CHI_I}
\end{equation}
at $x=1$.

With our compactified radial coordinate, and given the symmetries of our problem\footnote{From now on we consider only the cases with even parity with respect to $\theta=\pi/2$.}, a suitable spectral expansion for the black hole metric functions (collectively denoted by $\mathcal{F}=\{f,g,h,W\}$) is given by
\begin{equation}
  \mathcal{F}^{(k)} = \sum_{i=0}^{N_x-1} {\vphantom{\sum}}' \sum_{j=0}^{N_\theta-1} {\vphantom{\sum}}' \alpha_{ij}^{(k)} T_i(x) \cos \left(2j\theta\right),
  \label{eq:spectralexpansion}
\end{equation}
where $N_x$ and $N_{\theta}$ are the resolutions in the radial and angular coordinates. Note that, as discussed above,  the angular boundary conditions are automatically satisfied by this expansion (c.f. Table \ref{tab:trig}). 


As mentioned previously, we will usually use the Kerr metric itself to set our initial guess when working with modified theories of gravity, and to do so we will need the expression for the spectral coefficients that follow from an interpolation of a two-dimensional function $u(x,\theta)$, which is given by
\begin{equation}
  \alpha_{ij} = \frac{4}{N_x N_\theta} \sum_{k=0}^{N_x-1} \sum_{l=0}^{N_\theta-1} u(x_k,\theta_l) T_i(x_k) \cos\left(2j \theta_l\right),
  \label{eq:interpolation}
\end{equation}
where $x_k$ and $\theta_l$ are given in Eqs. \eqref{eq:chebnodes} and \eqref{eq:thetapoints} respectively.

\par
\begin{figure}[]
  \centering
      \includegraphics[width=0.65\textwidth]{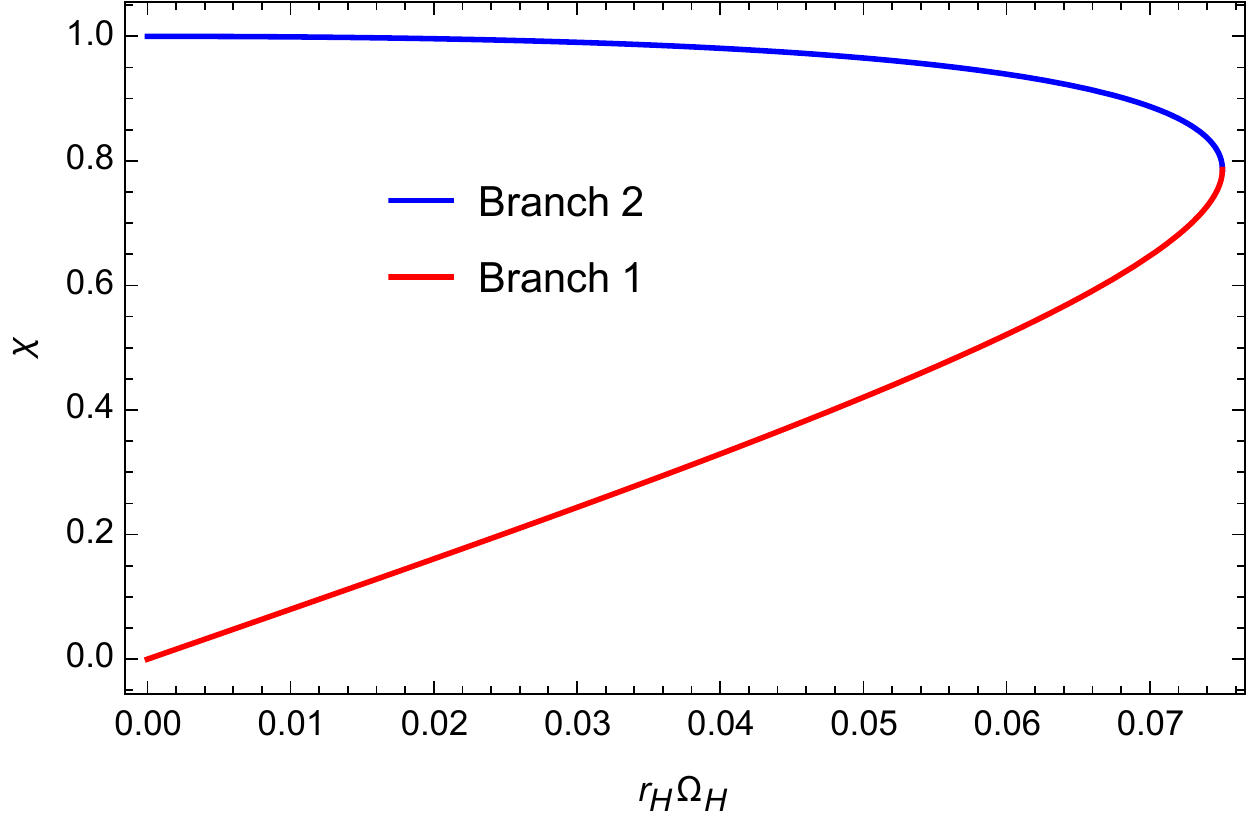}
  \caption{Choosing $(r_H, \Omega_H)$ as input parameters, for fixed $r_H$, two branches exist.}
  \label{fig:kerrbranches}
\end{figure}

 Each Kerr black hole is uniquely described by two input parameters. For example, in the presentation given in Eq. \eqref{eq:kerr}, these are the location of the event horizon $r_H$ and the ADM mass $M$. We have seen, however, in expressions \eqref{eq:rh} and \eqref{eq:omh} that they are related to the dimensionless spin $\chi$ and the horizon angular velocity $\Omega_H$. Therefore, using the correct parametrization, the Kerr solution can be described by any input pair chosen from $r_H$, $\Omega_H$, $\chi$, and $M$. In the numerical approach, in a theory agnostic setting, one input parameter that must be used is $r_H$ because it enters directly the metric ansatz and the definition of our compactified coordinate $x$. We have, however, freedom in the choice of the other input parameter in the numerics. To the best of our knowledge, so far in the literature for similar problems \cite{Herdeiro:2014goa, Herdeiro:2015gia, Herdeiro:2016tmi,Delgado:2020rev, Kleihaus:2015aje,Kleihaus:2011tg,Herdeiro:2020wei,Berti:2020kgk,Cunha:2019dwb}, the other input parameter has always been chosen as the event horizon angular velocity $\Omega_H$. Using this input pair $(r_H, \Omega_H)$, we find compatibility with the boundary conditions for the function $W$ if we choose Eqs. \eqref{eq:WBC_H_1} and \eqref{eq:W0_BC_x1} at the horizon and infinity, respectively. Then, in the case of a Kerr black hole, one finds that for a fixed value of $r_H$, two branches of solutions exist, as shown in Fig. \ref{fig:kerrbranches}. This follows from inverting the relation \eqref{eq:omh}.
The first branch of solutions starts at a vanishing value of $\Omega_H$ (for fixed $r_H$) and exists until 
\begin{equation}
  r_H \Omega_H = \frac{\sqrt{5\sqrt{5}-11}}{4\sqrt{2}} \approx 0.0750708,
\end{equation}
at which point
\begin{equation}
  \chi = \sqrt{\frac{\sqrt{5}-1}{2}} \approx 0.786151.
\end{equation}
Then, a second branch appears, and $\Omega_H$ tends backwards towards zero. As $\Omega_H \to 0$ on this second branch, extremal solutions are approached. The existence of two branches of solutions is not unique to Kerr, and is observed as well in the modified theories of gravity to be discussed in this work. We note that the numerical procedure gets rather difficult as near-extremal solutions are approached, as our metric ansatz with the described boundary conditions is not compatible with extremal solutions.

A novel approach that we can also adopt is to choose the pair $(r_H, \chi)$ as the input pair. This input pair is compatible with the $W$ boundary conditions of Eqs. \eqref{eq:W_BC_CHI_H} and \eqref{eq:W_BC_CHI_I} while maintaining the number of input parameters at two. We often find it very convenient to use the dimensionless spin as an input parameter, for example when  exploring domains of existence, or simply when working on a single solution where a certain $\chi$ is wanted. Our numerical spectral method is not only powerful because high accuracy solutions are produced, but also because highly non-linear boundary conditions can be handled with ease (which is the case of the boundary condition of Eq. \eqref{eq:W_BC_CHI_I}).

\begin{figure*}[t!]
  \centering
      \includegraphics[width=\textwidth]{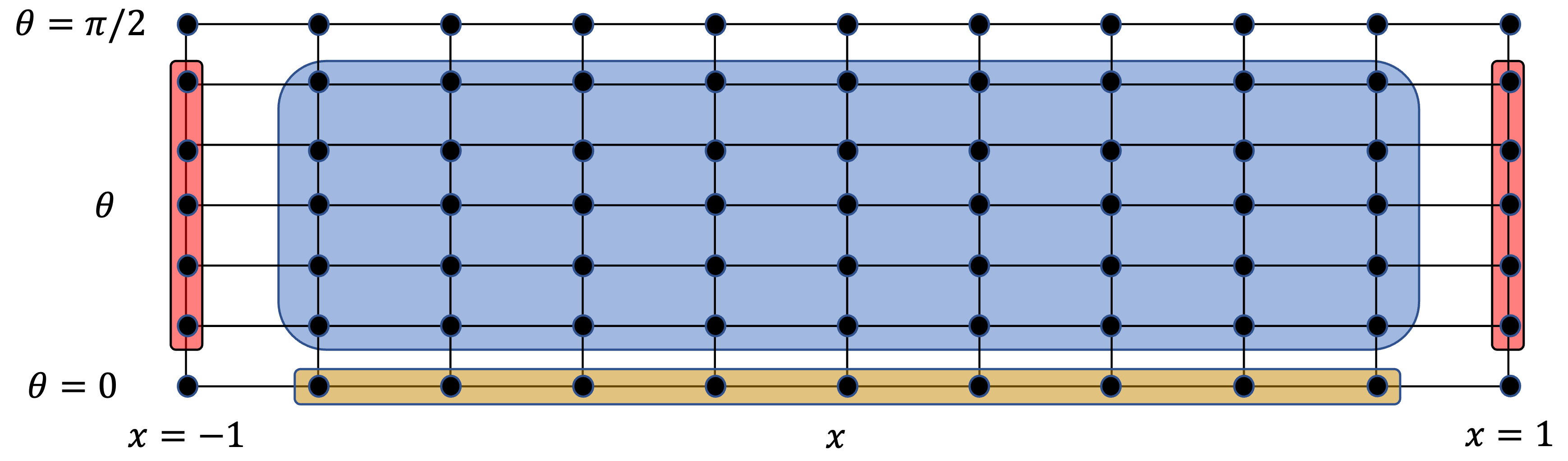}
  \caption{Fiducial grid with $N_x \times N_\theta = 11 \times 5$, highlighted in blue. The field equations (residuals) are evaluated in the blue region, with the boundary conditions imposed on the red region. The yellow highlight concerns the imposition of the condition of Eq. \eqref{eq:conicalsing}.}
  \label{fig:grid}
\end{figure*}

To solve the system of field equations subject to the discussed boundary conditions we must construct a suitable grid. This is done as follows. We assume a resolution $N_x \times N_\theta$. The discrete grid points in the $x$ direction are chosen according to Eq. \eqref{eq:chebnodes}, where we take $N = N_x-2$, together with the boundary points $x=- 1$ and $x=1$, such that the total number of points in the $x$ direction is $N_x$\footnote{This approach is also called \textit{boundary-bordering} in the spectral methods' literature.}. In $\theta$, our points are chosen as in Eq. \eqref{eq:thetapoints}, where we take $N=N_\theta$. The $x$ and $\theta$ points together form the  schematically shown in Fig. \eqref{fig:grid}, in blue.
Assuming there are a total number,  $N_{funcs}$, of functions to solve for, there are $N_{funcs}\times N_x \times N_\theta$ degrees of freedom (spectral coefficients) in the problem, as seen in the spectral expansion of Eq. \eqref{eq:spectralexpansion}. 
For each value of $\theta$ in the grid at the $x$ boundaries we impose for each function the horizon and asymptotic boundary conditions as discussed before. This gives us a total of $N_{funcs} \times 2 \times N_\theta$ equations (Fig. \ref{fig:grid}, in red). The remaining $N_{funcs} \times \left(N_x-2\right) \times N_\theta$ equations come from imposing the $N_{funcs}$ residuals resulting from the field equations at each non-boundary $x$ value, for each $\theta$. The number of degrees of freedom is then equal to the number of equations to solve, as it should. A small caveat -- the absence of conical singularities imposes that Eq. \eqref{eq:conicalsing} must be obeyed (i.e. for our coordinate range, $h=1$ at $\theta=0$). While we could leave this condition outside the numerical scheme and use it as another test to the code, we find that imposing it allows obtaining solutions with (much) higher accuracy. In our particular implementation, therefore, we have swapped the evaluation of one of the residuals at $\theta=0$ (for all interior values of $x$)\footnote{We empirically found that any of the field equations should equally valid to remove for this process, resulting in similar outcomes} with the condition of Eq. \eqref{eq:conicalsing}, see Fig. \ref{fig:grid} in yellow.

\subsubsection{Numerical Approach: A summary}
\par Here we summarize our numerical approach for clarity. To solve the field equations, some preliminary work must be done. First, we employ the metric ansatz of Eq. \eqref{eq:metric} which contains four unknown functions, $f$, $g$, $h$, and $W$. Plugging this metric ansatz onto the field equations of the theory, leads to a set of non-linear coupled PDEs that depend on the functions and their first and second derivatives $\left(\mathcal{F}, \partial_r \mathcal{F},\partial_r^2 \mathcal{F},\partial_\theta \mathcal{F},\partial_\theta^2 \mathcal{F}, \partial_{r\theta} \mathcal{F}\right)$. The set of field equations is then expressed in terms of the compactified coordinate $x$ defined in Eq. \eqref{eq:x} and put in residual form (i.e., $\mathcal{R}\left(x,\theta, \partial\mathcal{F}\right) = 0$). The same is done for the appropriate boundary conditions as discussed. This part of the process is usually done resorting to a computer algebra system such as \emph{Mathematica}, \emph{Maple} or \emph{SageMath}\footnote{In this work we have used \emph{Mathematica} along with the \emph{OGRe} package \cite{Shoshany:2021iuc} to obtain the explicit field equations of many theories.}. Our code, which can be found at \cite{gitlink}, includes detailed examples demonstrating how to derive the elliptic field equations for the two theories we will discuss: General Relativity and Einstein-Scalar-Gauss-Bonnet gravity. These examples are implemented using \emph{Mathematica}. They serve as a valuable reference and can be easily adapted to different contexts. Due to their complexity, these elliptic equations can consist of hundreds or even thousands of independent terms, hence we won't present them here. The residuals (and appropriate Jacobian) are then exported to a \emph{Julia} coding file in order to solve the problem using the developed numerical infrastructure. Each function is expanded in a spectral series given by Eq. \eqref{eq:spectralexpansion} and the input parameters are then specified (depending on the chosen boundary conditions for the function $W$). To successfully solve the field equations, a good initial guess must be provided to our Newton solver. For this, we interpolate the functions of the known Kerr solution using Eq. \eqref{eq:interpolation}, obtaining appropriate spectral coefficients to be provided as a good initial guess. If new fields are present, as is the case with modified theories, we typically take advantage of perturbative solutions and interpolate them as a guess. Convergence is assumed once the norm difference between the spectral coefficients of two successive iterations is less than a certain prescribed tolerance. 

To speed up the solver, the values of our basis functions and their first and second derivatives are calculated at all the grid points and stored, such that no repeated evaluations are performed. Another optimization that we found particularly impactful was to store the values on the grid of the trigonometric functions that typically appear in the residuals, $\sin \theta$ and $\cos \theta$.

Once a solution is obtained, physical quantities can be extracted from it as we discuss in the next section, and the solution can be used for numerous investigations.

\subsection{Physical Properties of Stationary and Axisymmetric Black Holes}
\label{sec:physical_properties}
\par Once a numerical  stationary and axisymmetric black hole solution has been found using our code, we can extract important quantities of physical relevance. In this section, we review many of the quantities that one can extract from a solution, some of which can be used to test the accuracy of our code. We have implemented additional code to extract all these quantities from a numerical solution.

\subsubsection{Quantities of interest}
Starting with the asymptotic quantities, we have seen that the mass $M$ and angular momentum $J$ can be extracted from the asymptotic expansion of Eq. \eqref{eq:asymptoticexpansionsmet} or Eq. \eqref{eq:asymptoticexpansions}. In terms of the coordinate $x$ these are given by
\begin{equation}
  M = r_H \left(1+\partial_x f\right)|_{x=1}, \qquad J = -r_H^2 \partial_x W|_{x=1}.
\end{equation}
We remark that such a simple expression for the extraction of $J$ is the reason why we have defined the function $W$ in this way -- such that its decay is of the form $\sim 1/r$, allowing for more accurate results. In a circular spacetime, the zeroth law of black hole mechanics holds, which means that the surface gravity is constant on the horizon of the stationary black hole. The surface gravity is defined as $\kappa^2 = -1/2 (\nabla_\mu \xi_\nu)(\nabla^\mu \xi ^\nu)$, where $\xi$ was defined in Eq. \eqref{eq:killingvector1}. The Hawking temperature \cite{Hawking:1975vcx} can then be obtained from the surface gravity as
\begin{equation}
  T_H = \frac{\kappa}{2\pi} = \left. \frac{1}{2\pi r_H} \frac{f}{\sqrt{g h}} \right|_{x=-1}.
\end{equation}
The induced metric on the horizon is
\begin{equation}
  d\Sigma^2 = \mathfrak{h}_{ij} dx^i dx^j = r_H^2 \left.\frac{g}{f}\left[h d\theta^2 + \sin^2 \theta d\varphi^2\right] \right|_{x=-1},
\end{equation}
and from it we can compute several quantities of interest, the most important being the event horizon area
\begin{equation}
  A_H = \int_H \sqrt{\mathfrak{h}} d\theta d\varphi = \left. 2\pi r_H^2 \int_{0}^{\pi} d\theta \sin \theta \frac{g \sqrt{h}}{f} \right|_{x=-1}.
\end{equation}
Also of importance is the entropy, which is given in the Iyer-Wald formalism by \cite{Iyer:1994ys}
\begin{equation}
  S = \left. -2\pi \int_H \frac{\delta \mathcal{L}}{\delta R_{\mu \nu \alpha \beta}} \epsilon_{\mu \nu} \epsilon_{\alpha \beta} dA \right|_{\mathrm{on-shell}},
  \label{eq:entropy_wald}
\end{equation}
where $\epsilon_{\mu \nu}$ is the binormal vector to the event horizon surface. In the case of a Kerr black hole the above expression reduces to the simple form $S = A_H/4$.
The horizon and asymptotic quantities are connected via the Smarr type relation \cite{PhysRevLett.30.71,Bardeen:1973gs,Iyer:1994ys,Liberati:2015xcp}
\begin{equation}
  M = \left. 2T_H S + 2 \Omega_H J - 2 \int_{\Sigma} d^3x \sqrt{-g} \mathcal{L}\right|_{\mathrm{on-shell}}.
  \label{eq:smarr}
\end{equation}
The Smarr relation is extremely important when studying numerical solutions as it provides a test to the code that relates physical quantities obtained on the horizon and asymptotic regions, allowing us to estimate the accuracy of the numerical method.
Also of interest is the perimetral radius $\mathfrak{R}$ which is a geometrically significant radial coordinate such that a circumference along the equatorial plane has perimeter $2\pi \mathfrak{R}$. It is related to the coordinate $r$ by
\begin{equation}
  \mathfrak{R} = \left. \sqrt{g_{\phi \phi}} \right|_{\theta = \pi/2} = \left. \sqrt{\frac{g}{f}} r \right|_{\theta = \pi/2}.
\end{equation}
To explore the horizon geometry, it is useful to define the horizon circumference along the equator
\begin{equation}
  L_e = 2\pi \mathfrak{R}_H,
\end{equation}
and along the poles
\begin{equation}
  L_p = 2\int_{0}^{\pi} \sqrt{g_{\theta \theta}}|_{x=-1}d\theta = \left. 2r_H \int_{0}^{\pi} \sqrt{\frac{gh}{f}}  \right|_{x=-1} d\theta.
\end{equation}
With these two quantities, we can define the sphericity
\begin{equation}
  s = \frac{L_e}{L_p}.
\end{equation}
For a Kerr black hole $s \geq 1$, with $s$ increasing with spin. That means that spin deforms the horizon towards oblateness. The linear velocity of the horizon quantifies how fast the null geodesic generators of the horizon spin relative to a static observer at infinity, and is given by
\begin{equation}
  v_H = \Omega_H \mathfrak{R}_H.
\end{equation}

For a Kerr black hole we have in terms of $M$ and $r_H$
\begin{equation}
  \begin{aligned}
    &J = M^2\sqrt{1-\left(\frac{2r_H}{M}\right)^2}, \\&
    T_H = \frac{1}{4\pi M \left(1+\frac{M}{2r_H}\right)},\\&
    A_H = 8\pi M^2\left(1+\frac{2r_H}{M}\right) \\&
    L_e = 4\pi M, \\&
    L_p = 4M\sqrt{2\left(1+\frac{2r_H}{M}\right)} \mathrm{EllipticE}\left(\frac{1}{2}\left[1-\frac{2r_H}{M}\right]\right), \\&
    \mathfrak{R}_H = 2M,
  \end{aligned}
\end{equation}
where $\mathrm{EllipticE}$ denotes the complete elliptic integral of the second kind, and we also note that $2r_H/M = \sqrt{1-\chi^2}$.
The Kerr solution is Ricci flat, and thus the Lagrangian of GR vanishes on-shell and therefore so does the last term in Eq. \eqref{eq:smarr}.

\subsubsection{Ergoregion}
\par
\begin{figure}[]
  \centering
      \includegraphics[width=0.75\textwidth]{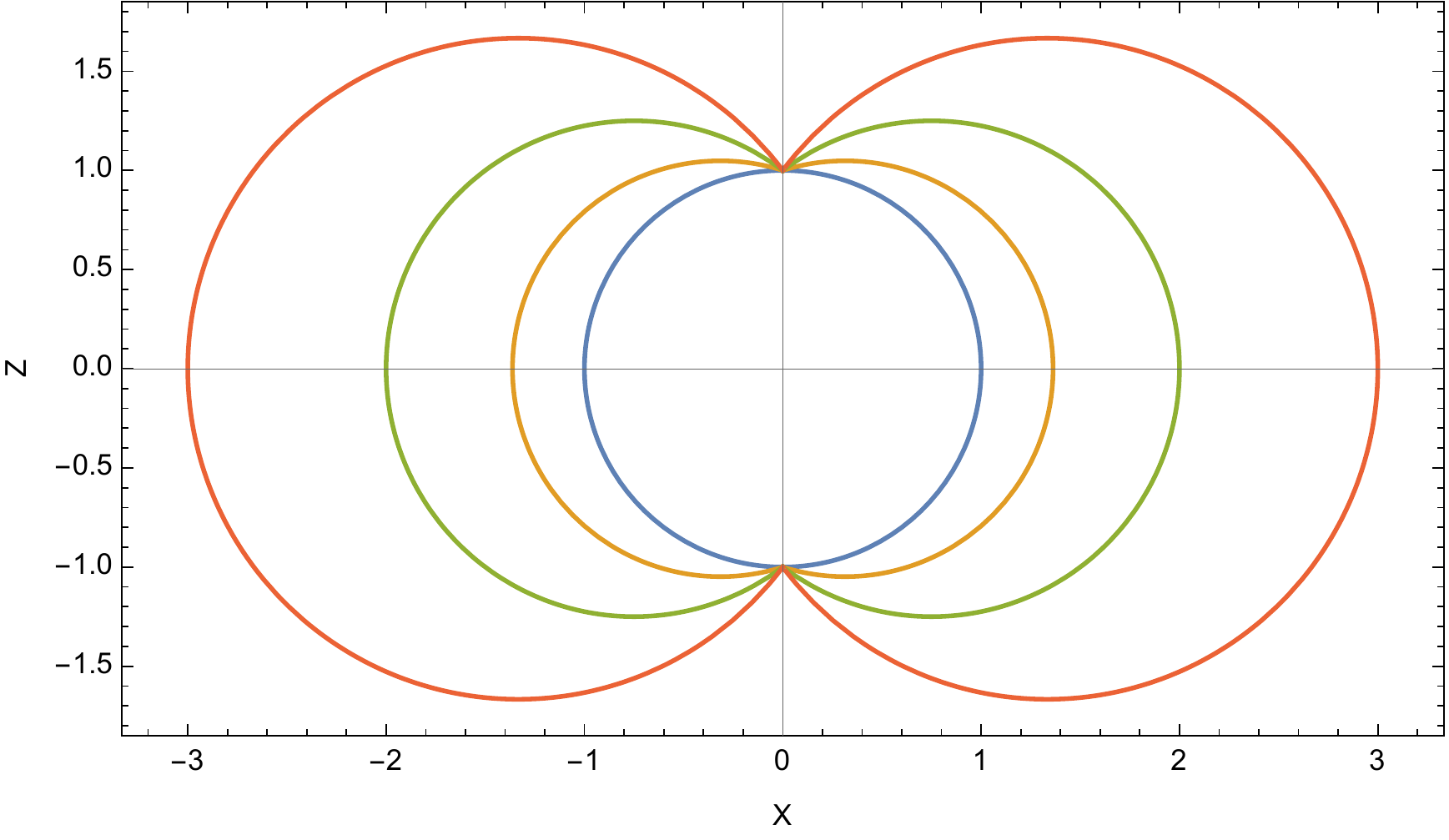}
  \caption{Ergoregion of a Kerr black hole with $\chi = 0.3$ (orange), $\chi = 0.6$ (green), and $\chi = 0.8$ (red) visualized on the $X-Z$ plane. The event horizon is shown in blue.}
  \label{fig:ergokerr}
\end{figure}

 The ergoregion is defined as the domain outside the event horizon wherein the norm of the asymptotically timelike Killing vector $k=\partial_t$ becomes positive, $g_{\mu \nu} k^\mu k^\nu > 0$. It is bounded by the event horizon and by the surface where
\begin{equation}
  g_{tt} = -f \mathcal{N}^2 + \frac{g \left(1-\mathcal{N}\right)^2 W^2}{f} \sin^2\theta = 0.
\end{equation}
Within the ergoregion, an object cannot appear stationary with respect to a distant observer due to the intense frame-dragging.\footnote{This immediately follows from the fact that the 4-velocity of a massive particle must be timelike, $g_{\mu \nu} u^\mu u^\nu<0$. Indeed, the worldline of an object standing still at a fixed point implies that $u = \partial_t$, and if $g_{tt}\geq 0$, then $g_{\mu \nu} u^\mu u^\nu\geq 0$.} Furthermore, ergoregions raise the possibility of extracting energy from a black hole via the Penrose process, or superradiant scattering \cite{Brito:2015oca}. Starting from the well-known result for the ergosphere of a Kerr black hole in Boyer-Lindquist coordinates and inverting the relation of Eq. \eqref{eq:coord_transf} we obtain that in quasi-isotropic coordinates the ergosphere of a Kerr black hole is located at
\begin{equation}
  r_E^{Kerr} = \frac{r_H}{\sqrt{1-\chi^2}} \left(\sqrt{1-\chi^2 \cos^2\theta}+\chi \sin \theta\right),
\end{equation}
where the subscript ``$E$" refers to ``ergoregion".
Due to the symmetries of our problem, we need only consider the range $\theta \in [0,\pi/2]$. To visualize the ergoregion, we introduce the coordinates
\begin{equation}
  X = \frac{r}{r_H} \sin \theta, \qquad Z = \frac{r}{r_H} \cos \theta.
\end{equation}
In Fig. \ref{fig:ergokerr} we observe the ergoregion of a Kerr black holes in the $X-Z$ plane for several values of dimensionless spin.

\subsubsection{Petrov type}
\par The Petrov classification allows for a kinematic characterization of the gravitational field in a coordinate independent manner using algebraic properties of the Weyl tensor $C_{\mu \nu \alpha \beta}$, namely its number of distinct principal null directions. This classification is useful, for example, when searching for exact solutions, or for a Carter-like constant \cite{Carter:1968}. Using the Newman-Penrose formalism, the information is contained in five complex scalars known as the Weyl scalars. With the null tetrad $\{l^\mu, n^\mu, m^\mu, \overline{m}^\mu\}$, where $l^{\mu}$ and $n^{\mu}$ are real, and $m^\mu, \overline{m}^\mu$ are complex conjugate satisfying the orthonormality conditions $l^\mu n_\mu=1$, $m^\mu \overline{m}_\mu=-1$ and all other products zero, the Weyl scalars are defined as 
\begin{equation}
    \begin{aligned}
        &\psi_0 = -C_{\mu \nu \alpha \beta} l^\mu m^\nu l^\alpha m^\beta, \\&
        \psi_1 = -C_{\mu \nu \alpha \beta} l^\mu n^\nu l^\alpha m^\beta, \\&
        \psi_2 = -C_{\mu \nu \alpha \beta} l^\mu m^\nu \overline{m}^\alpha n^\beta, \\&
        \psi_3 = -C_{\mu \nu \alpha \beta} l^\mu n^\nu \overline{m}^\alpha n^\beta, \\&
        \psi_4 = -C_{\mu \nu \alpha \beta} n^\mu \overline{m}^\nu n^\alpha \overline{m}^\beta.
    \end{aligned}
\end{equation}
With the above scalars, the following Lorentz invariant quantities can be constructed
\begin{equation}
    \begin{aligned}
        &I = \psi_0 \psi_4 - 4\psi_1 \psi_3 + 3 \psi_2^2, \\&
        J = -\psi_2^3 + \psi_0 \psi_2 \psi_4 + 2\psi_1 \psi_2 \psi_3 - \psi_4 \psi_1^2 - \psi_0 \psi_3^2,\\&
        D = I^3 - 27 J^2,\\&
        K = \psi_4^2 \psi_1 - 3\psi_4 \psi_3 \psi_2 + 2\psi_3^3,\\&
        L = \psi_4 \psi_2 - \psi_3 ^2,\\& 
        N = 12L^2-\psi_4^2 I.
    \end{aligned}
\end{equation}
Given the above quantities, it is possible to determine the Petrov type of a given spacetime. The classification is summarized in Table \ref{tab:petrov} \cite{Achour:2021pla}. In particular, a spacetime is said to be \textit{algebraically special} if $D = 0$. The Kerr(-Newman) spacetime is Petrov type D. In a numerical setup, we also find useful to introduce the \textit{speciality index} defined as \cite{Berti:2004ny}
\begin{equation}
    S = \frac{27 J^2}{I^3}.
    \label{eq:specindex}
\end{equation}
\begin{table}[]
  \centering
  \begin{tabular}{|c|c|}
  \hline
  Type & Conditions \\ \hline
  O & $\psi_0 = \psi_1 = \psi_2 = \psi_3 = \psi_4 = 0$ \\ \hline
  I & $D \neq 0$ \\ \hline
  II & $D = 0, I\neq 0, J \neq 0, K \neq 0, N \neq 0$ \\ \hline
  III & $D = 0, I = J = 0, K \neq 0, L \neq 0$ \\ \hline
  N & $D = 0, I = J = K = L = 0$ \\ \hline
  D & $D = 0, I \neq 0,  J \neq 0, K = N = 0$ \\ \hline
  \end{tabular}
  \caption{Summary of Petrov classification.}
  \label{tab:petrov}
\end{table}
With an appropriate choice of tetrad, following Ref. \cite{Berti:2004ny}, it is possible to gauge away $\psi_1$ and $\psi_3$ to zero. Such a tetrad would be for example
\begin{equation}
    \begin{aligned}
        &l^\mu = \sqrt{\frac{g_{\varphi \varphi}}{2 \left(g_{t\varphi}^2 - g_{tt} g_{\varphi \varphi}\right)}} \left(1,0,0,-\frac{g_{t\varphi}+\sqrt{g_{t\varphi}^2 - g_{tt} g_{\varphi \varphi}}}{g_{\varphi \varphi}}\right),\\&
        n^\mu = \sqrt{\frac{g_{\varphi \varphi}}{2 \left(g_{t\varphi}^2 - g_{tt} g_{\varphi \varphi}\right)}} \left(1,0,0,-\frac{g_{t\varphi}-\sqrt{g_{t\varphi}^2 - g_{tt} g_{\varphi \varphi}}}{g_{\varphi \varphi}}\right),\\&
        m^\mu = \frac{1}{\sqrt{2}} \left(0,\frac{i}{\sqrt{g_{rr}}},\frac{1}{\sqrt{g_{\theta \theta}}},0\right).
    \end{aligned}
\end{equation}

\subsubsection{Marginal Stable Circular Orbits: Light Rings and ISCO}
\par The study of marginal stable circular orbits is highly relevant for the observational properties of black holes. The innermost stable circular orbit (ISCO) of massive particles is the smallest possible radius for a stable circular orbit and is often taken to mark the inner edge of an accretion disk around a black hole. Accelerated charged particles orbiting the black hole emit synchroton radiation whose physical properties are connected with the frequency of geodesics at the ISCO. Therefore, physical properties of an astrophysical black hole can be inferred via measurements of the ISCO through accretion disks.
\par Light rings are circular null geodesics, typically unstable, allowing light to encircle a black hole before being scattered to infinity or falling into the event horizon. From an observational point of view, they are important for observations made with the Event Horizon Telescope as they are intimately connected with the shadow of the black hole \cite{Cunha:2018acu}.

To compute the ISCO and light rings we follow Ref. \cite{Sullivan:2020zpf}. We start by considering the line element of Eq. \eqref{eq:metric} in the form
\begin{equation}
  ds^2 = g_{tt} dt^2 + g_{rr} dr^2 + g_{\theta \theta} d\theta^2 + g_{\varphi \varphi} d\varphi^2 + 2g_{t \varphi} dt d\varphi.
\end{equation}
The two independent killing vectors of the spacetime, $k^\mu = \partial_t$ and $\Phi^\mu = \partial_\varphi$, have the associated conserved reduced energy $E$ and angular momentum $L$
\begin{equation}
  \begin{aligned}
    &E = - k_\mu \frac{dx^\mu}{d\lambda} = - g_{tt} \dot t - g_{t \varphi} \dot \varphi,\\&
    L = \Phi_\mu \frac{dx^\mu}{d\lambda} = g_{t \varphi} \dot t + g_{\varphi \varphi} \dot \varphi,
  \end{aligned}
\end{equation}
where $\dot \equiv d/d\lambda$. The above expressions can be rearranged in terms of $\dot t$ and $\dot \varphi$
\begin{equation}
  \begin{aligned}
    &\dot t = \frac{E g_{\varphi \varphi} + L g_{t \varphi}}{g_{t \varphi}^2 - g_{tt}g_{\varphi \varphi}},\\&
    \dot \varphi = - \frac{E g_{t \varphi} + L g_{tt}}{g_{t \varphi}^2 - g_{tt}g_{\varphi \varphi}}.
  \end{aligned}
  \label{eq:dottdotp}
\end{equation}
Considering orbits restricted to the equatorial plane, $\theta = \pi/2$, the condition associated with the normalization of the four-velocity of the particles becomes
\begin{equation}
  -\epsilon = g_{tt} \dot t^2 + g_{rr} \dot r^2 + g_{\varphi \varphi} \varphi^2 +2 g_{t \varphi} \dot t \dot \varphi,
\end{equation}
with $\epsilon = \{0,1,-1\}$ for a massless, massive and tachyon particle, respectively. We disregard $\epsilon=-1$ from now on. Substituting the expressions of Eq. \eqref{eq:dottdotp} in the above condition, and solving for $\dot r^2$, we can define the effective potential
\begin{equation}
  U_{eff} = \frac{1}{g_{rr}} \left(-\epsilon + \frac{E^2 g_{\varphi \varphi} + 2E L g_{t\varphi} + L^2 g_{tt}}{g_{t \varphi}^2 - g_{tt}g_{\varphi \varphi}}\right),
\end{equation}
such that
\begin{equation}
  \dot r^2 = U_{eff}.
\end{equation}
The conditions for a circular orbit are $\dot r = 0$ and $\ddot r = 0$, from which follows that
\begin{equation}
  U_{eff} = 0, \qquad \frac{dU_{eff}}{dr} \equiv U_{eff}' = 0,
\end{equation}
at the location of orbit. The dash denotes a derivative with respect to $r$. These conditions can further be rearranged into algebraic equations that must be satisfied simultaneously
\begin{equation}
  \begin{aligned}
    &E^2 g_{\varphi \varphi} + 2E L g_{t\varphi} + L^2 g_{tt} - \epsilon \left(g_{t \varphi}^2 - g_{tt}g_{\varphi \varphi}\right) = 0,\\&
    E^2 g_{\varphi \varphi}' + 2E L g_{t\varphi}' + L^2 g_{tt}' - \epsilon \left(g_{t \varphi}^2 - g_{tt}g_{\varphi \varphi}\right)' = 0.
  \end{aligned}
  \label{eq:circular_orbits1}
\end{equation}

\noindent \textbf{Light Rings}\\
For a light particle, $\epsilon=0$. In this case, calculations are simpler than in the massive case. Solving the first equation for $L$ in \eqref{eq:circular_orbits1} and substituting in the second we obtain
\begin{equation}
  g_{\varphi \varphi}' + 2 g_{t \varphi}' \left(\frac{g_{t\varphi} \pm \sqrt{g_{t \varphi}^2 - g_{tt}g_{\varphi \varphi}}}{g_{tt}}\right) + g_{tt}' \left(\frac{g_{t\varphi} \pm \sqrt{g_{t \varphi}^2 - g_{tt}g_{\varphi \varphi}}}{g_{tt}}\right)^2 = 0,
\end{equation}
which is to be evaluated on a radius $r$. The smallest root of the above equation is the location of the light ring.

In Boyer-Lindquist coordinates the location of the circular photon orbits of a Kerr black hole are given by \cite{Rezzolla:2016}
\begin{equation}
  r_{BL}^{LR \pm} = 2M \left(1+\cos\left(\frac{2}{3}\arccos\left(\mp \chi\right)\right) \right),
\end{equation}
where the plus sign refers to co-rotating photons, and the minus sign to counter-rotating photons. In quasi-isotropic coordinates the location of the circular photon orbits can be obtained using the inverse transformation in Eq. \eqref{eq:coord_transf_inv}.

\noindent \textbf{ISCO}\\
For a massive particle, $\epsilon=1$. The ISCO is located at a saddle point of the effective potential, such that the condition $U_{eff}''=0$ should be imposed. This is equivalent to imposing
\begin{equation}
  E^2 g_{\varphi \varphi}'' + 2E L g_{t\varphi}'' + L^2 g_{tt}'' - \epsilon \left(g_{t \varphi}^2 - g_{tt}g_{\varphi \varphi}\right)'' = 0,
  \label{eq:circular_orbits2}
\end{equation}
in addition to Eq. \eqref{eq:circular_orbits1}. To find the location of the ISCO, we first solve Eq. \eqref{eq:circular_orbits1} for $E$ and $L$ as functions of the metric functions and their first derivatives, and later substitute these onto Eq. \eqref{eq:circular_orbits2}. Similarly to the light-ring case, we obtain a second order equation to be solved for $r$, the smallest root of which corresponds to the location of the ISCO.

In Boyer-Lindquist coordinates the location of the circular massive particle orbits of a Kerr black hole are given by \cite{Rezzolla:2016}
\begin{equation}
  r_{BL}^{ISCO\pm} = M \left(3+Z_2 \mp \sqrt{(3-Z_1)(3+Z_1+2Z_2)}\right),
\end{equation}
where
\begin{equation*}
  \begin{aligned}
    &Z_1 = 1+\left(1-\chi^2\right)^{1/3} \left[(1+\chi)^{1/3} + (1-\chi)^{1/3}\right],\\&
    Z_2 = \sqrt{3\chi^2+Z_1^2},
  \end{aligned}
\end{equation*}
and the plus sign refers to co-rotating particles, and the minus sign to counter-rotating particles. In quasi-isotropic coordinates the location of the circular orbits can be obtained using the inverse transformation in Eq. \eqref{eq:coord_transf_inv}.

\noindent \textbf{Orbital frequencies at the ISCO and Light Ring}\\
The orbital angular frequency of particles both at the ISCO and light ring is given by
\begin{equation}
    \omega_\pm = \frac{\dot \varphi}{\dot t} = \frac{-g_{t\varphi}' \pm \sqrt{g_{t\varphi}'^2 - g_{tt}' g_{\varphi \varphi}'}}{g_{\varphi \varphi}'},
\end{equation}
where the above expression is to be evaluated at the location of the ISCO/light ring, $\omega_+$ is the angular frequency of co-rotating particles and $\omega_-$ is the angular frequency of counter-rotating particles. In the case of a Kerr black hole we have
\begin{equation}
    M \omega_\pm = \pm \frac{1}{\sqrt{48 \cos^4 \left(\frac{1}{3}\arccos{\left(\mp \chi\right)}\right)+\chi^2}},
\end{equation}
at the light ring, and
\begin{equation}
    M \omega_\pm = \pm \frac{1}{\left(r_{BL}^{ISCO\pm}/M\right)^{3/2} \pm \chi},
\end{equation}
at the ISCO. The orbital frequency at the ISCO is associated with the cut-off frequency of the emitted synchrotron radiation generated from accelerated charges in accretion disks, and the angular frequency at the light ring is related to the time-scale of the response of the black hole when it is perturbed (real part of the frequency of the black hole quasi-normal modes) \cite{Cardoso:2008bp}.

\section{Numerical spinning black hole solutions}
\label{sec:numbhs}
In this section we first validate our numerical infrastructure against well-known results, namely the Kerr black hole, and then proceed to use it to obtain spinning black holes in a modified gravity theory, the Einstein-scalar-Gauss-Bonnet theory.

\subsection{Validating the code against the Kerr black hole}
\begin{figure}[]
  \centering
      \includegraphics[width=0.5\textwidth]{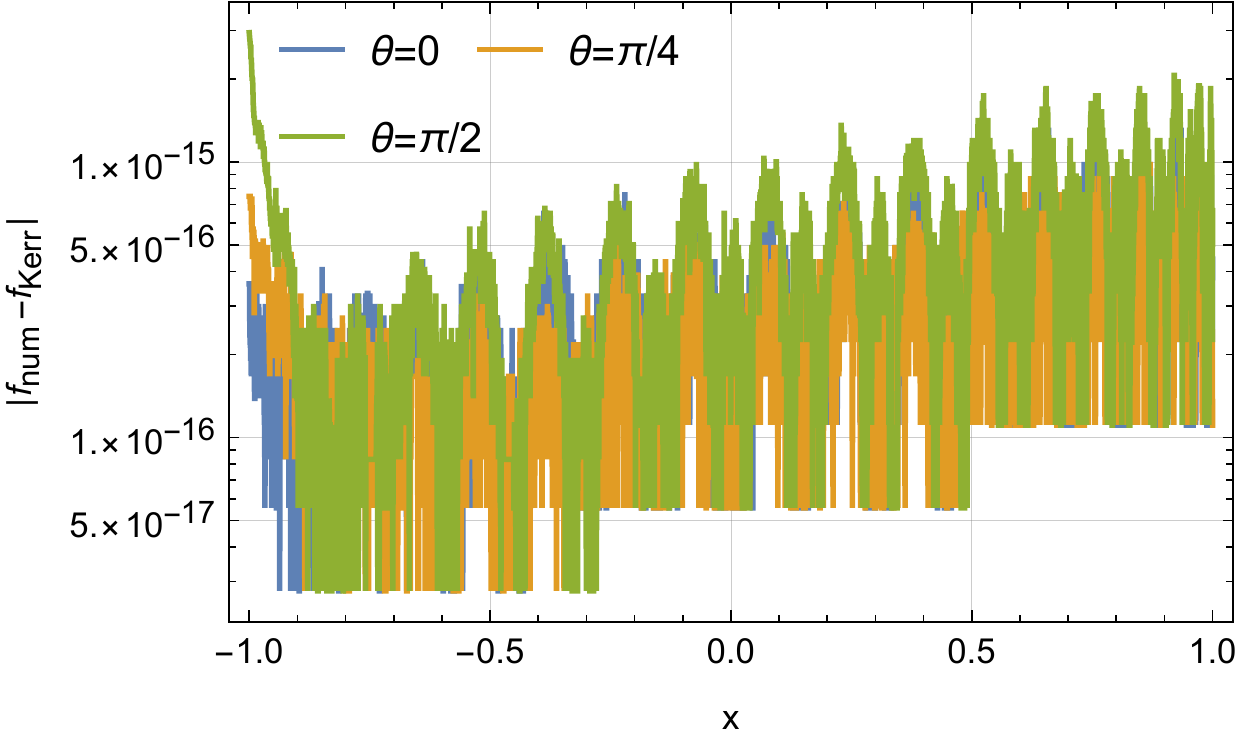}\hfill
      \includegraphics[width=0.5\textwidth]{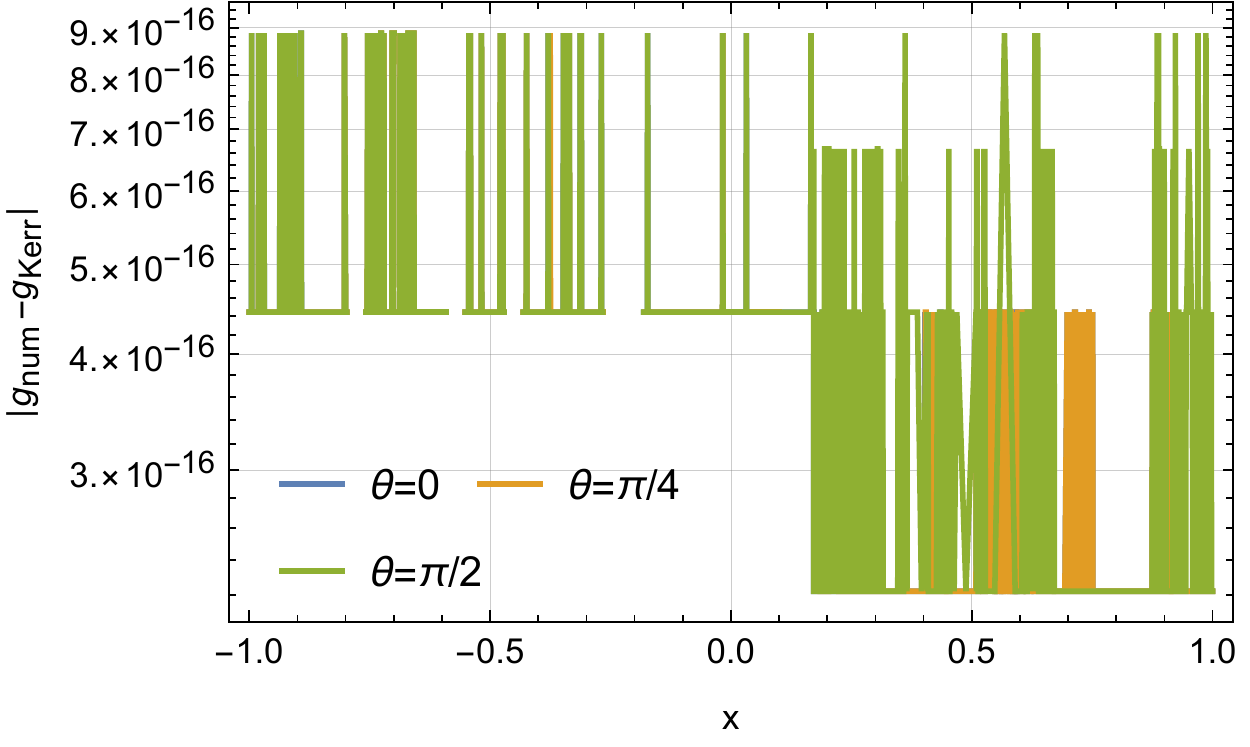}\vfill
      \includegraphics[width=0.5\textwidth]{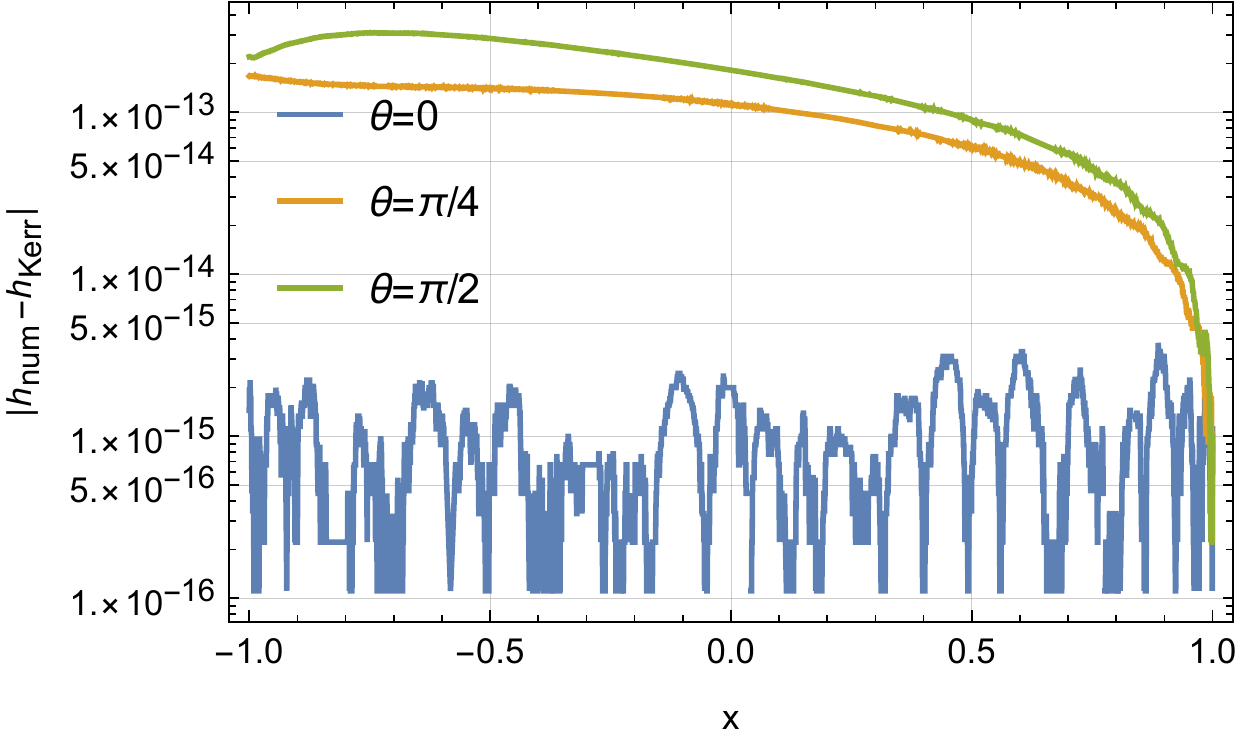}\hfill
      \includegraphics[width=0.5\textwidth]{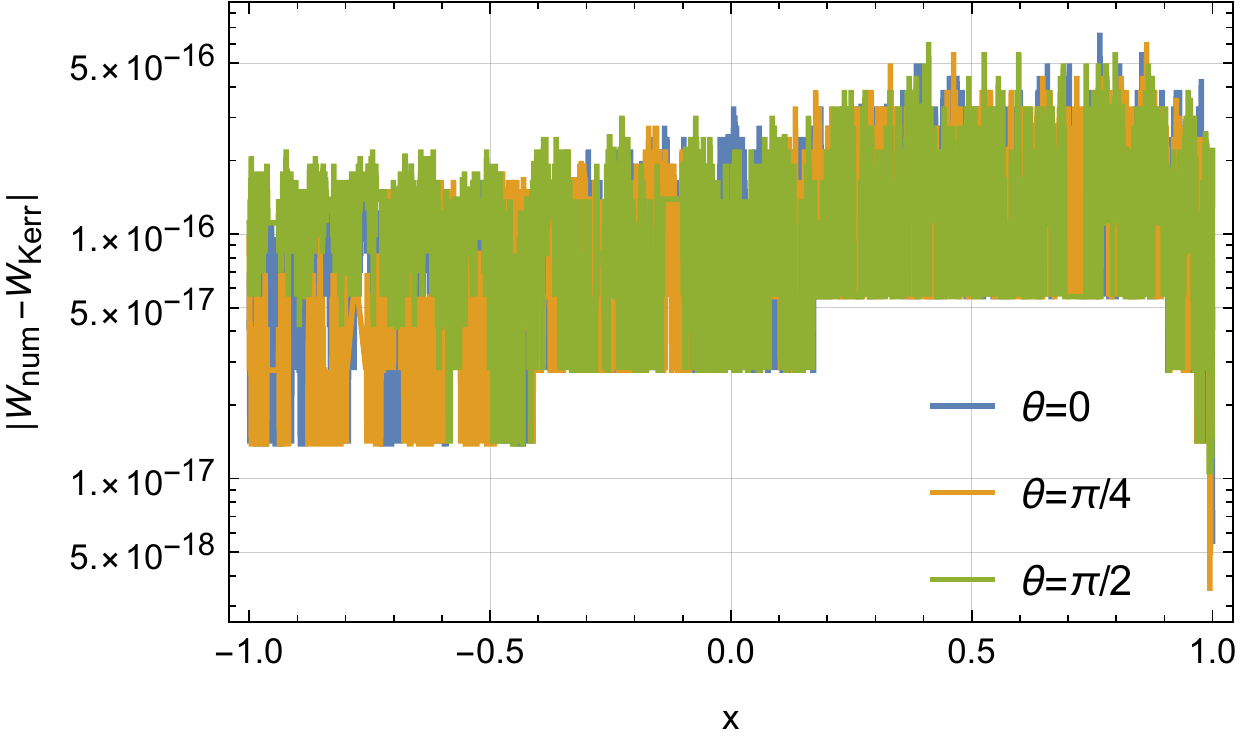}
  \caption{Comparison between the numerical and analytical results for a Kerr black hole with $\chi = 0.6$, using $N_x=42$, $N_\theta = 8$. The maximum observed error is of order $\mathcal{O}\left(10^{-13}\right)$ for the function $h$, with all other functions being obtained to machine precision. A Schwarzschild black hole was used as an initial guess, and we have used $r_H=1$.}
  \label{fig:compare_kerr_sol}
\end{figure}

To validate our numerical infrastructure we will solve the axisymmetric vacuum Einstein equations to numerically obtain the Kerr solution, and compare with analytical results. We choose to solve the the following combination of the field equations
which diagonalize the Einstein tensor with respect to the operator $\partial_r^2 + r^{-2}\partial_\theta^2$:
\begin{equation}
  \begin{aligned}
    &-\mathcal{E}^{\mu}_{\phantom{\mu} \mu} + 2 \mathcal{E}^{t}_{\phantom{t} t} + \frac{2W r_H}{r^2} \mathcal{E}^{\varphi}_{\phantom{\varphi} t} = 0,\\&
    \mathcal{E}^{\varphi}_{\phantom{\varphi} t} = 0,\\&
    \mathcal{E}^{r}_{\phantom{r} r} + \mathcal{E}^{\theta}_{\phantom{\theta} \theta} = 0,\\&
    \mathcal{E}^{\varphi}_{\phantom{\varphi} \varphi} - \frac{W r_H}{r^2} \mathcal{E}^{\varphi}_{\phantom{\varphi} t} - \mathcal{E}^{r}_{\phantom{r} r} - \mathcal{E}^{\theta}_{\phantom{\theta} \theta} = 0.
  \end{aligned}
  \label{eq:feqs_combination}
\end{equation}
In Fig. \ref{fig:compare_kerr_sol} we present the results for the comparison of the metric functions obtained numerically with the analytically known ones for a Kerr black hole with $\chi=0.6$. Given that in this case the initial guess cannot be the Kerr metric itself, to obtain the results in Fig. \ref{fig:compare_kerr_sol} we used a  Schwarzschild black hole with comparable $r_H$\footnote{We find the code to be robust against initial guesses, converging quickly even when these are somewhat (but not extremely) bad.}. The maximum observed error is of $\mathcal{O}\left(10^{-13}\right)$ for the metric function $h$, with all other metric functions being successfully obtained to machine precision.
We also explored the whole domain of existence of Kerr black holes, comparing numerically obtained physically relevant quantities with analytical ones, see Fig. \ref{fig:kerr_domain} below. These include the mass $M$, angular momentum $J$, horizon area $A_H$ and Hawking temperature $T_H$ of the black holes. Furthermore, we computed the (normalized) Smarr relation in Eq. \eqref{eq:smarr}. Overall, in all quantities we have found remarkable agreement between numerical and analytical results, with the Smarr relation providing accurate maximum error estimates. We also observe that errors are higher when the black holes approach the extremal case $(\chi \to 1)$. This is because in the extremal limit, our setup is not valid and another metric \"ansatz is needed (see e.g. Ref. \cite{Herdeiro:2015gia}). 
\begin{figure}[]
    \centering
        \includegraphics[width=0.5\textwidth]{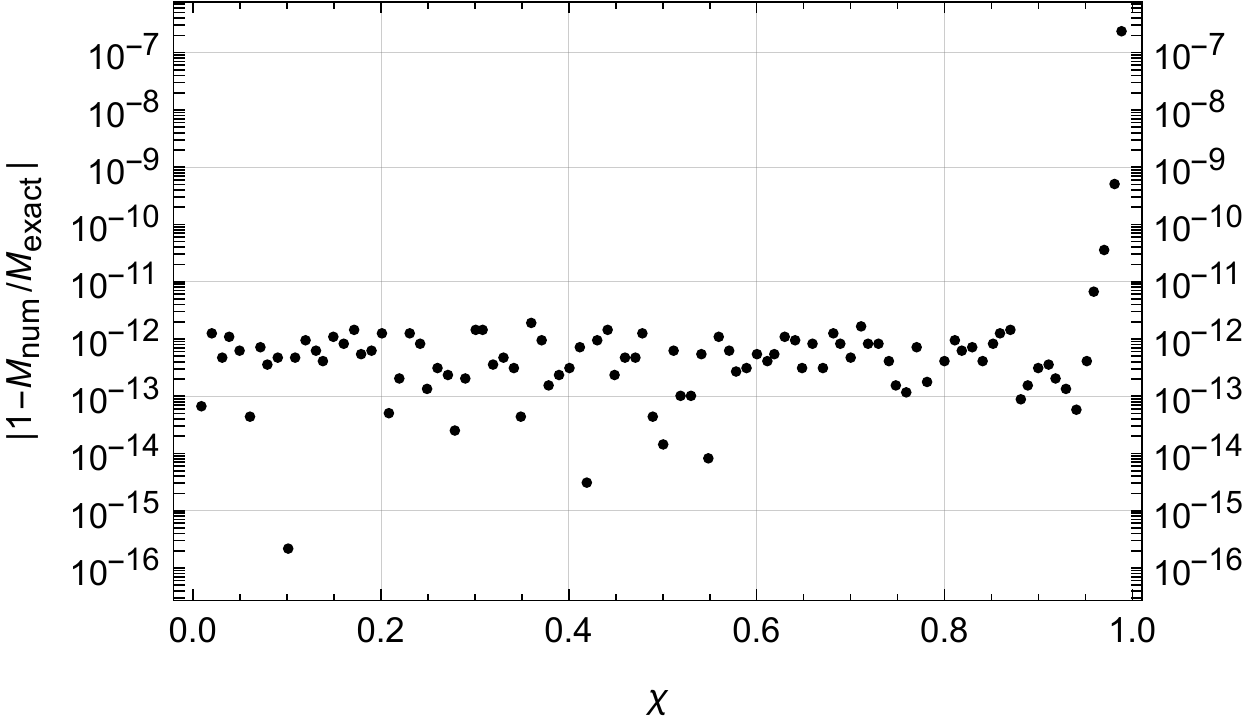}\hfill
        \includegraphics[width=0.5\textwidth]{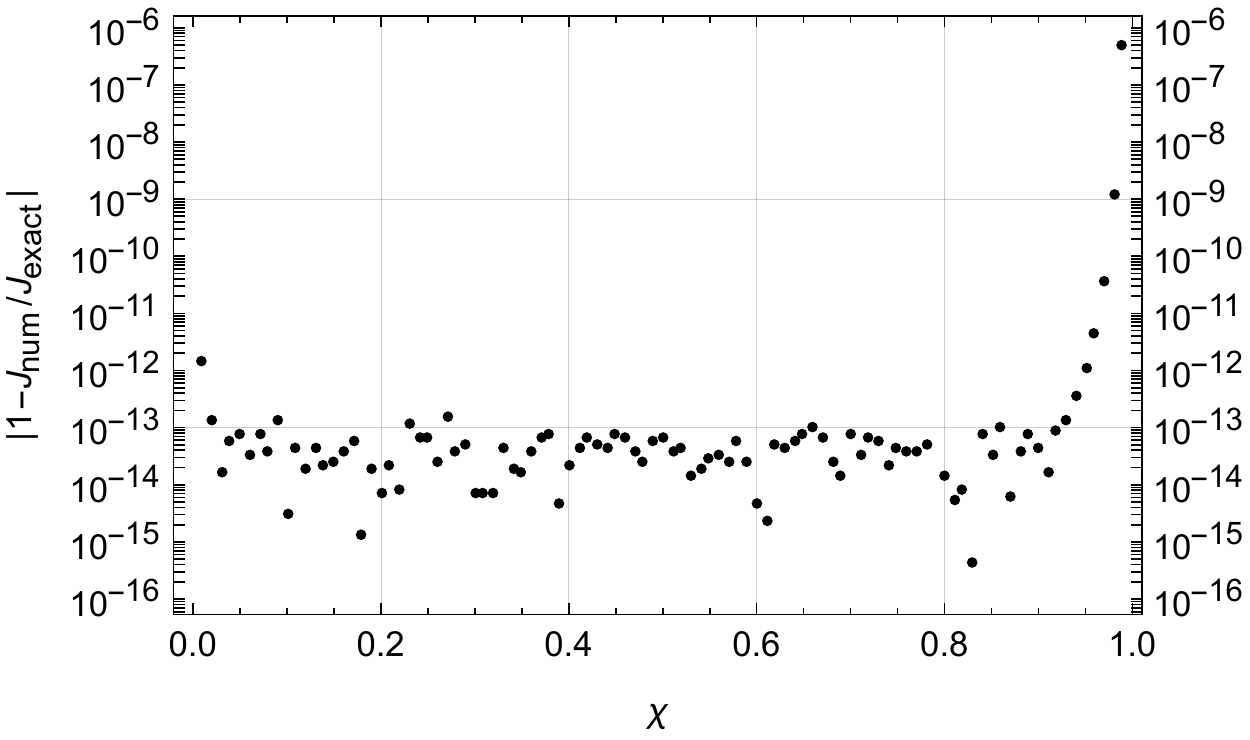}\vfill
        \includegraphics[width=0.5\textwidth]{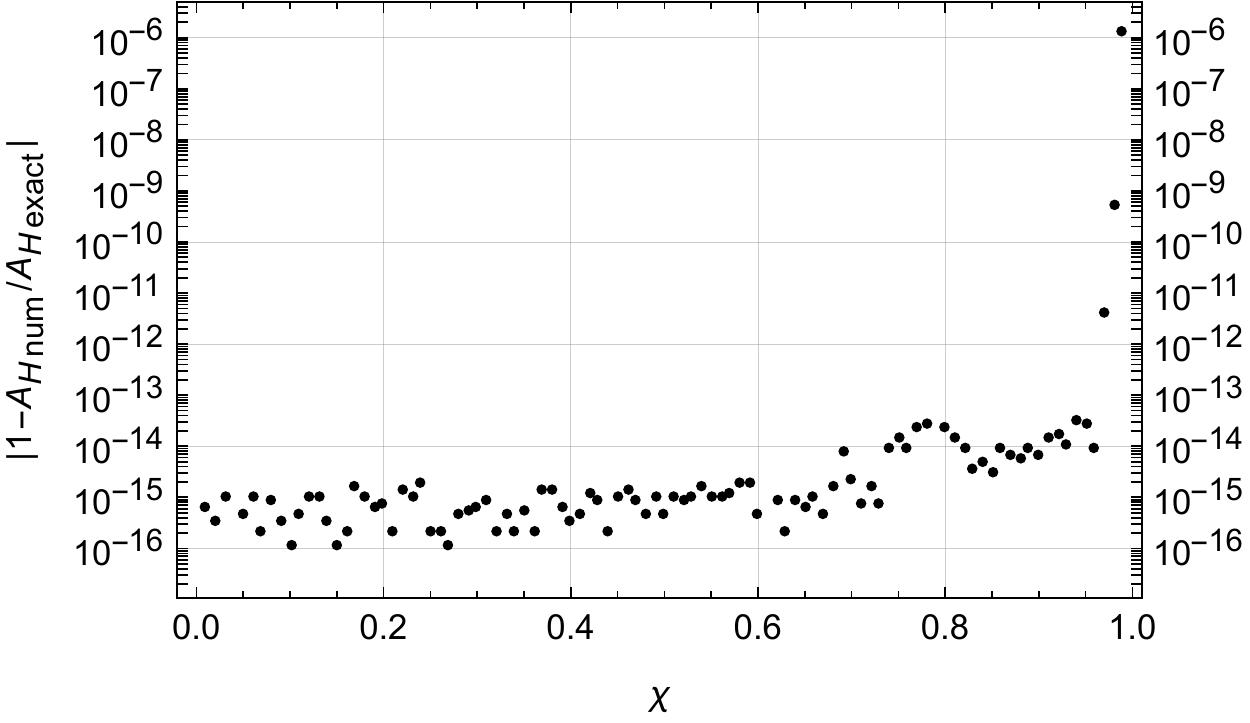}\hfill
        \includegraphics[width=0.5\textwidth]{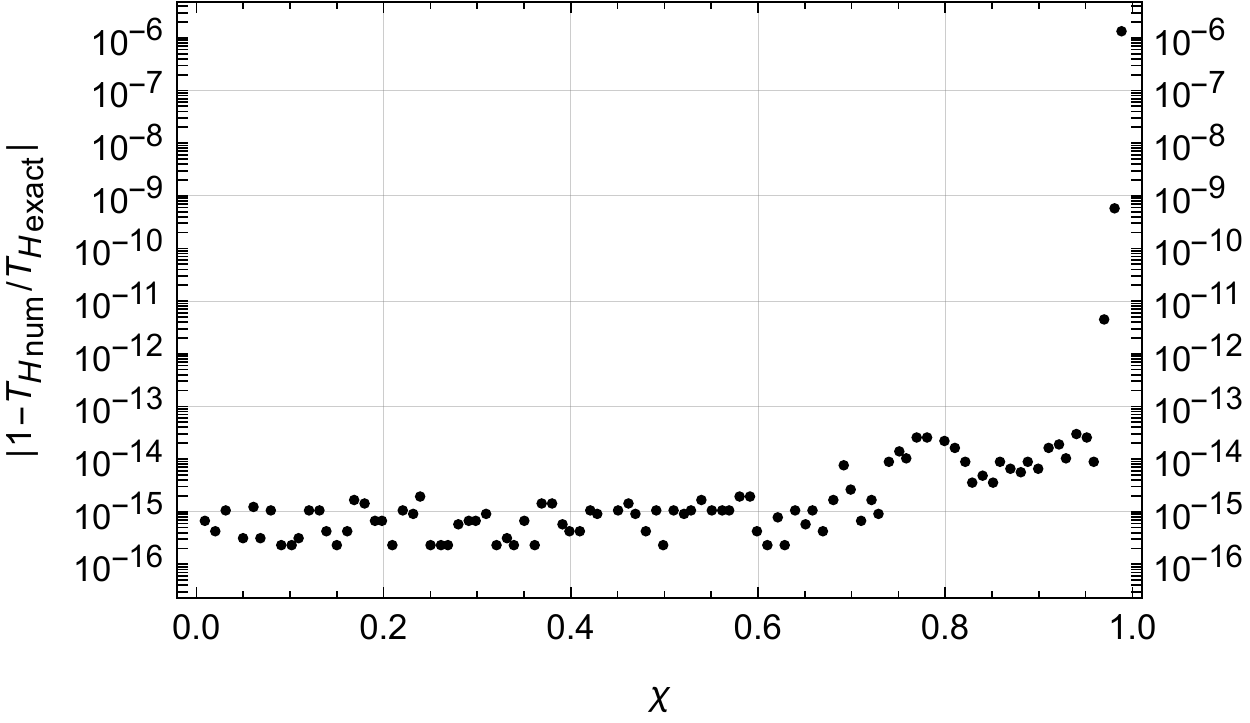}\vfill
        \includegraphics[width=0.5\textwidth]{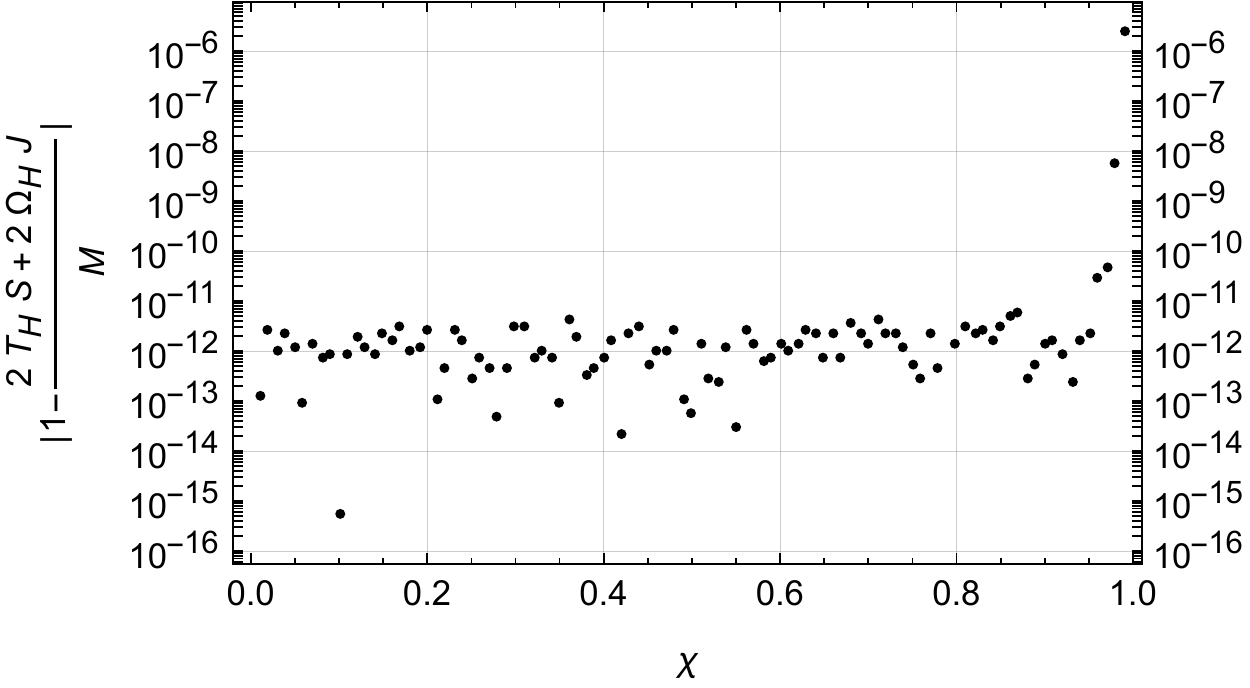}\vfill
    \caption{Comparison of numerical results for $M$, $J$, $A_H$ and $T_H$ with analytical ones, throughout the domain of existence of Kerr black holes. Each point represents a different black hole solution. Numerical results were obtained using $N_x=50$, $N_\theta = 12$. We observe remarkable agreement and small errors overall.}
    \label{fig:kerr_domain}
\end{figure}

\subsection{Einstein-scalar-Gauss-Bonnet Gravity}
Einstein-scalar-Gauss-Bonnet (EsGB) theories of gravity are a popular set of scalar tensor theories of gravity that have been extensively studied \cite{Sotiriou:2013qea,Sotiriou:2014pfa,Doneva:2017bvd,Silva:2017uqg,Antoniou:2017acq,Cunha:2019dwb,Dima:2020yac,Herdeiro:2020wei,Berti:2020kgk,Kanti:1995vq,Kleihaus:2011tg,Cunha:2016wzk,Delgado:2020rev}, and which admit black hole solutions different to those of GR. Here we use this set of theories to test our methods and code in a non-trivial, but previously studied setting.
EsGB theories are described by the action
\begin{equation}
    \mathcal{S} = \frac{1}{16\pi} \int d^4x \sqrt{-g}\left(R - \left(\nabla \phi\right)^2 + \frac{\alpha}{4}\xi\left(\phi\right) \mathcal{G} \right),
    \label{eq:EsGB6}
\end{equation}
where  $\phi$ is a real scalar field that couples non-minimally to the Gauss-Bonnet term via the coupling function $\xi(\phi)$, and where $\alpha$ is a coupling constant with dimensions of length squared. No closed-form black hole solutions are known in these models, even in the static case. One is therefore forced to resort to numerical methods to study black holes in these theories.

The field equations of the action \eqref{eq:EsGB6} are
\begin{equation}
  \mathcal{E}_{\mu \nu} \equiv G_{\mu \nu} - T_{\mu \nu} = 0,
  \end{equation}
  where
  \begin{equation*}
      T_{\mu \nu} = \nabla_\mu \phi \nabla_\nu \phi - \frac{1}{2} g_{\mu \nu} \dpp + \alpha\, P_{\mu \alpha \nu \beta} \nabla^\alpha \nabla^\beta \xi\cbr{\phi},
  \end{equation*}
  and
  \begin{equation*}
          P_{\alpha \beta \mu \nu} \equiv \frac{1}{4} \epsilon_{\alpha \beta \gamma \delta} R^{\rho \sigma \gamma \delta} \epsilon_{\rho \sigma \mu \nu} = 2\, g_{\alpha [\mu}G_{\nu] \beta} + 2\, g_{\beta [\nu} R_{\mu] \alpha} -R_{\alpha \beta \mu \nu},
  \end{equation*}
  is the double-dual Riemann tensor (the square brackets denote anti-symmetrization). The scalar field equation is
  \begin{equation}
  \mathcal{E}_{\phi} \equiv \boxp + \frac{\alpha}{8} \dot \xi(\phi) \GB = 0,
  \label{eq:sfequation_num}
  \end{equation}
where the dot denotes differentiation with respect to the scalar field $\phi$. In the stationary and axisymmetric setting, we find that the scalar field is subject to the boundary conditions \cite{Sullivan:2019vyi,Sullivan:2020zpf}
\begin{equation}
  \begin{aligned}
    &\partial_x \phi = 0, \qquad x=-1,\\&
    \phi = 0, \qquad x=1,\\&
    \partial_\theta \phi = 0, \qquad \theta = 0,\pi/2,
  \end{aligned}
\end{equation}
while the boundary conditions for the metric functions remain those given above. We therefore choose the same spectral expansion for the scalar field as we did for the metric functions.

Black holes in the EsGB theory should obey the Smarr formula \eqref{eq:smarr}, which becomes
\begin{equation}
    M + M_s = 2T_H S + 2 \Omega_H J,
\end{equation}
where\footnote{This relation can also be written as $$ M_s = \frac{1}{4\pi} \int d^3x \sqrt{-g} \left(\nabla \phi\right)^2 \frac{\partial}{\partial \phi} \left(\frac{\xi(\phi)}{\xi'(\phi)} \right),$$ provided the coupling does not obey $\xi(\phi) \propto \xi'(\phi)$ and the scalar field asymptotically vanishes. This is advantageous from a numerical point of view because no second derivatives of the scalar field are required, increasing the accuracy in computing $M_s$.}
\begin{equation}
    M_s = -\frac{1}{4\pi} \int d^3x \sqrt{-g} \frac{\xi(\phi)}{\xi'(\phi)} \Box \phi,
\end{equation}
and the entropy is given by Eq. \eqref{eq:entropy_wald} that in the EsGB case becomes
\begin{equation}
  S = \frac{A_H}{4} + \frac{\alpha}{8}\int_H d^2x \sqrt{\mathfrak{h}} \xi(\phi) R^{(2)},
\end{equation}
where $R^{(2)}$ is the Ricci scalar of the induced metric on the horizon.
We will focus on two coupling examples, the linear coupling
\begin{equation}
  \xi(\phi) = \phi,
\end{equation}
and the exponential coupling
\begin{equation}
  \xi(\phi) = e^{\gamma \phi}.
\end{equation}
We find that for the exponential coupling the Smarr relation takes a rather simple form
\begin{equation}
  M + Q_s/\gamma = 2T_H S + 2 \Omega_H J,
  \label{eq:smarr_esgb_exp}
\end{equation}
where $Q_s$ is the scalar charge of the solution, appearing in the asymptotic expansion of the scalar field
\begin{equation*}
  \phi \approx \frac{Q_s}{r} + \mathcal{O}\left(r^{-2}\right).
\end{equation*}
It can also be proved that for the linear coupling the following relation holds \cite{Prabhu:2018aun}
\begin{equation}
  Q_s = 2\pi \alpha T_H.
  \label{eq:rel_esgb_lin}
\end{equation}

\begin{figure}[]
  \centering
      \includegraphics[width=0.6\textwidth]{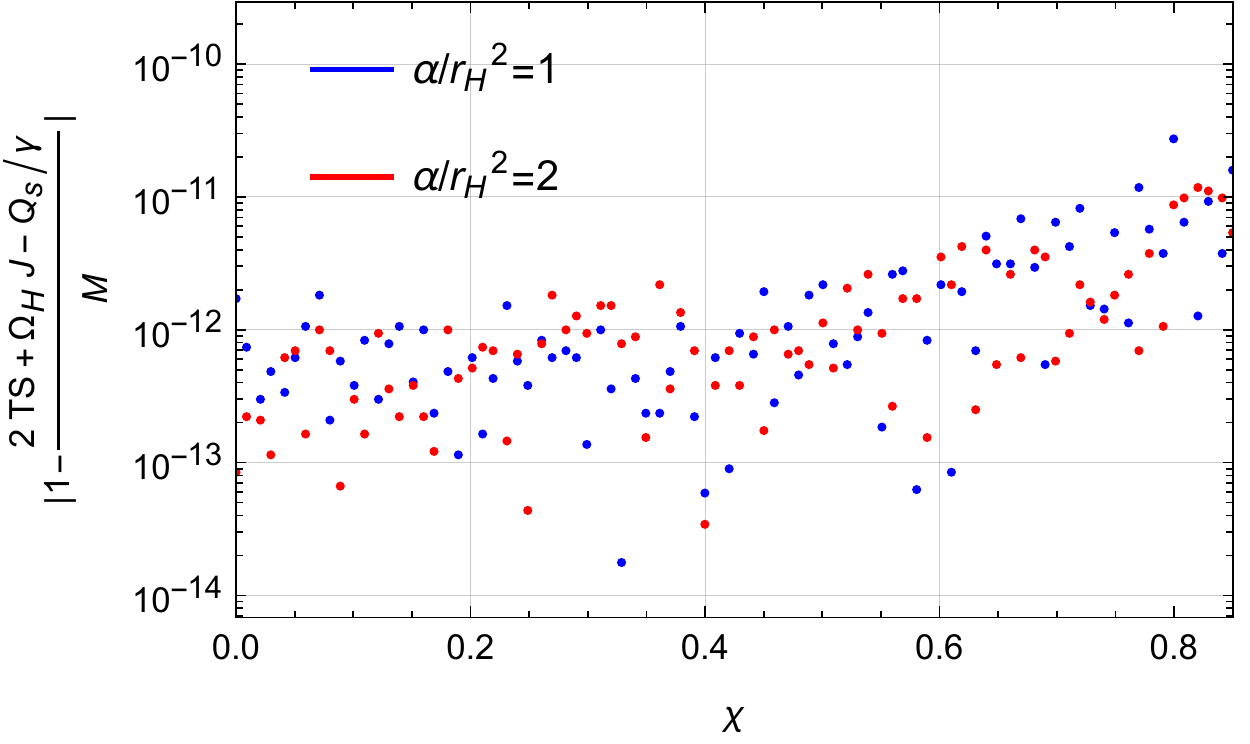}\vfill
      \includegraphics[width=0.6\textwidth]{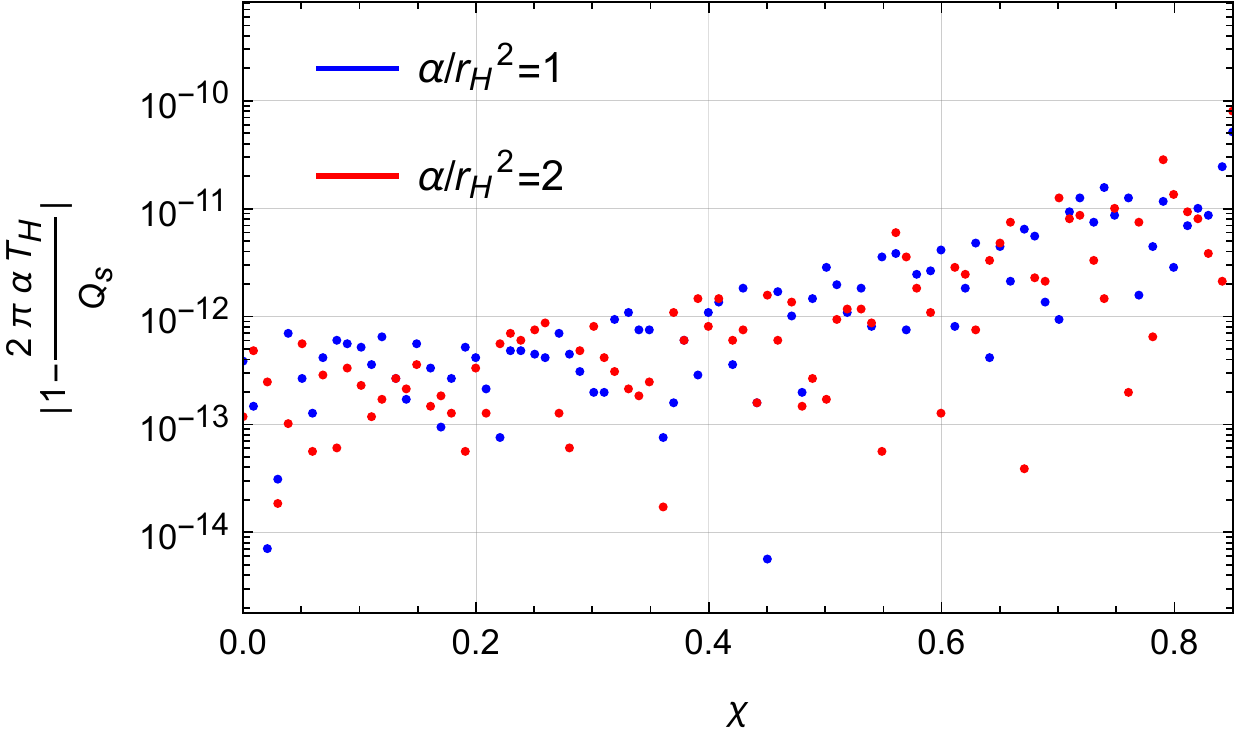}
  \caption{Smarr relation (top) and relation in Eq. \eqref{eq:rel_esgb_lin} (bottom) for numerical solutions in a part of the domain of existence for the theory with the exponential coupling with $\gamma=1$ and linear coupling, respectively, for different values of $\alpha/r_H^2$. Each point represents a different black hole solution. Numerical results were obtained using $N_x=50$, $N_\theta = 12$. We observe small errors, similarly to the Kerr case.}
  \label{fig:EsGB_domain_example_exp}
\end{figure}

In what follows we use the relations in Eqs. \eqref{eq:smarr_esgb_exp} and \eqref{eq:rel_esgb_lin} to address the accuracy of our numerical solutions for the exponential and linear couplings respectively. This is necessary as closed-form solutions are unknown. We use the same combination of field equations as in the Kerr case (Eq. \eqref{eq:feqs_combination}), along with the scalar field equation \eqref{eq:sfequation_num} to solve the system. To solve the system we use a comparable Kerr black hole as an initial guess for the metric functions, and for the scalar field we use the perturbative solution \cite{Sullivan:2019vyi,Sullivan:2020zpf}
\begin{equation}
    \phi \approx \frac{\alpha}{r_H^2} \frac{415-1047 x+942 x^2-358 x^3+51 x^4-3 x^5}{12 (-3+x)^6}.
\end{equation}
We present the accuracy estimate results (in a part of the domain of existence) using the Smarr relation for the exponential coupling and the relation in Eq. \eqref{eq:rel_esgb_lin} for the linear coupling in Fig. \ref{fig:EsGB_domain_example_exp}.
We observe that errors, as measured by the relations \eqref{eq:smarr_esgb_exp} and \eqref{eq:rel_esgb_lin}, are small and similar to those presented for the Kerr black hole in Fig. \ref{fig:kerr_domain}, despite a dramatic increase in the complexity and number of terms in the field equations. Our results also agree remarkably well with perturbative solutions, such as the ones obtained in Ref. \cite{Sullivan:2020zpf}.

\begin{figure}[]
  \centering
      \includegraphics[width=0.5\textwidth]{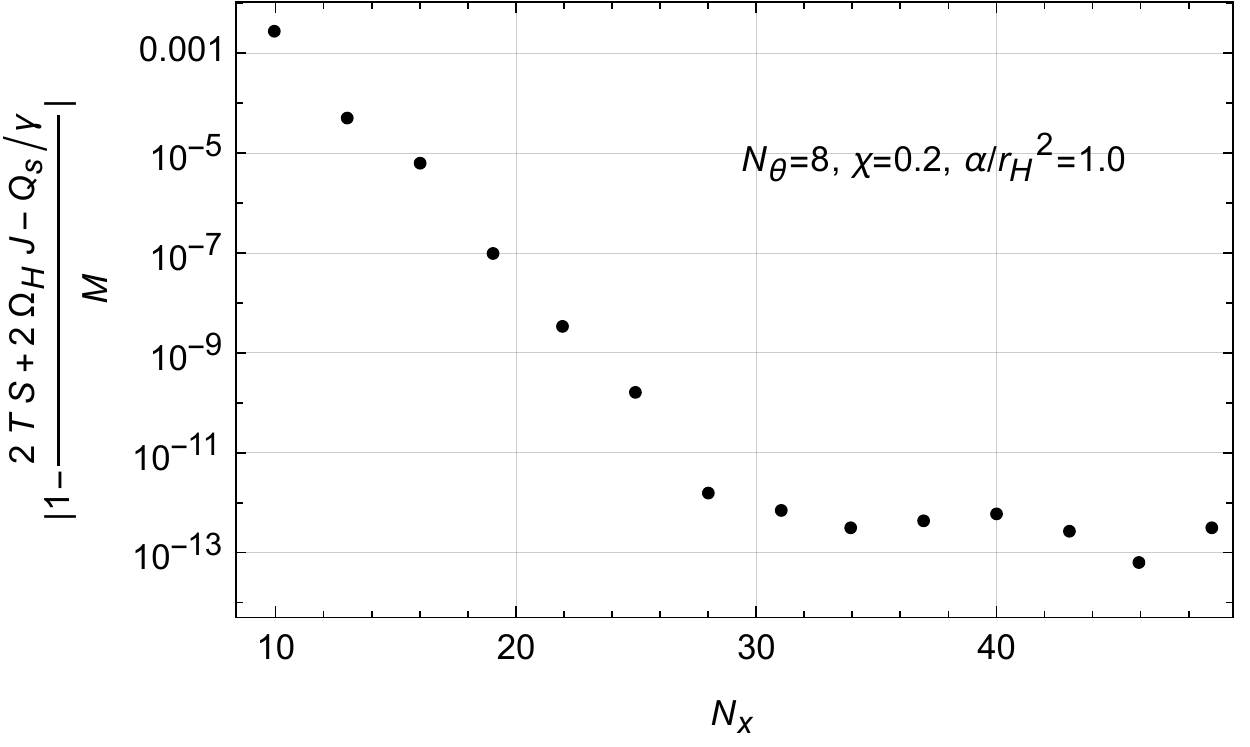}\hfill
      \includegraphics[width=0.5\textwidth]{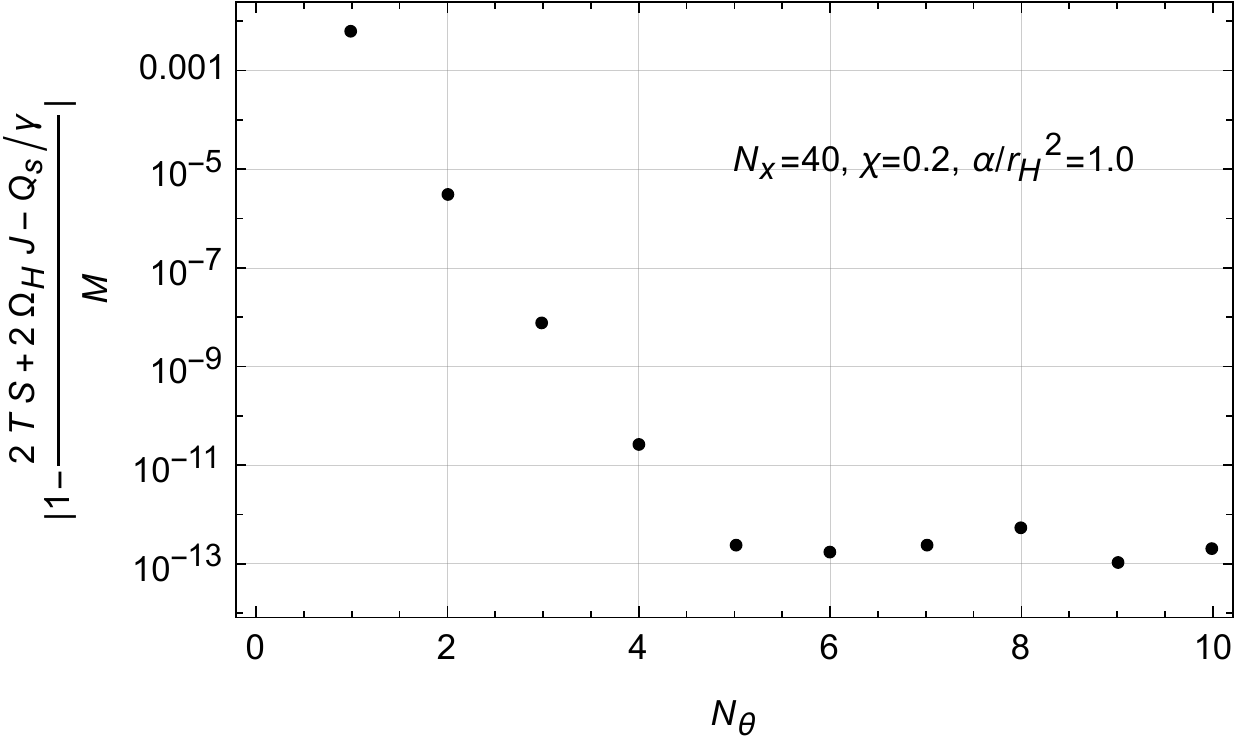}
  \caption{Smarr relation for numerical solutions with a dilaton coupling ($\gamma=1$) as a function of the resolution in $x$ (left) and $\theta$ (right). We observe exponential convergence to as the resolution is increased.}
  \label{fig:EsGB_VaryN}
\end{figure}

As another test to the code, in Figure \ref{fig:EsGB_VaryN} we plot the accuracy as estimated by the Smarr relation \eqref{eq:smarr_esgb_exp} as a function of both resolutions $N_x$ and $N_\theta$. We observe exponential convergence, similarly to the toy model presented in Fig. \ref{fig:odeerror}. Note that the Smarr relation provides only an estimate of maximum error -- recall the Kerr case, where most metric functions were actually obtained to a precision of $\sim \mathcal{O}\left(10^{-16}\right)$ but the Smarr relation attained errors on the order of $\sim \mathcal{O}\left(10^{-13}\right)$.

\begin{figure}[]
  \centering
      \includegraphics[width=0.5\textwidth]{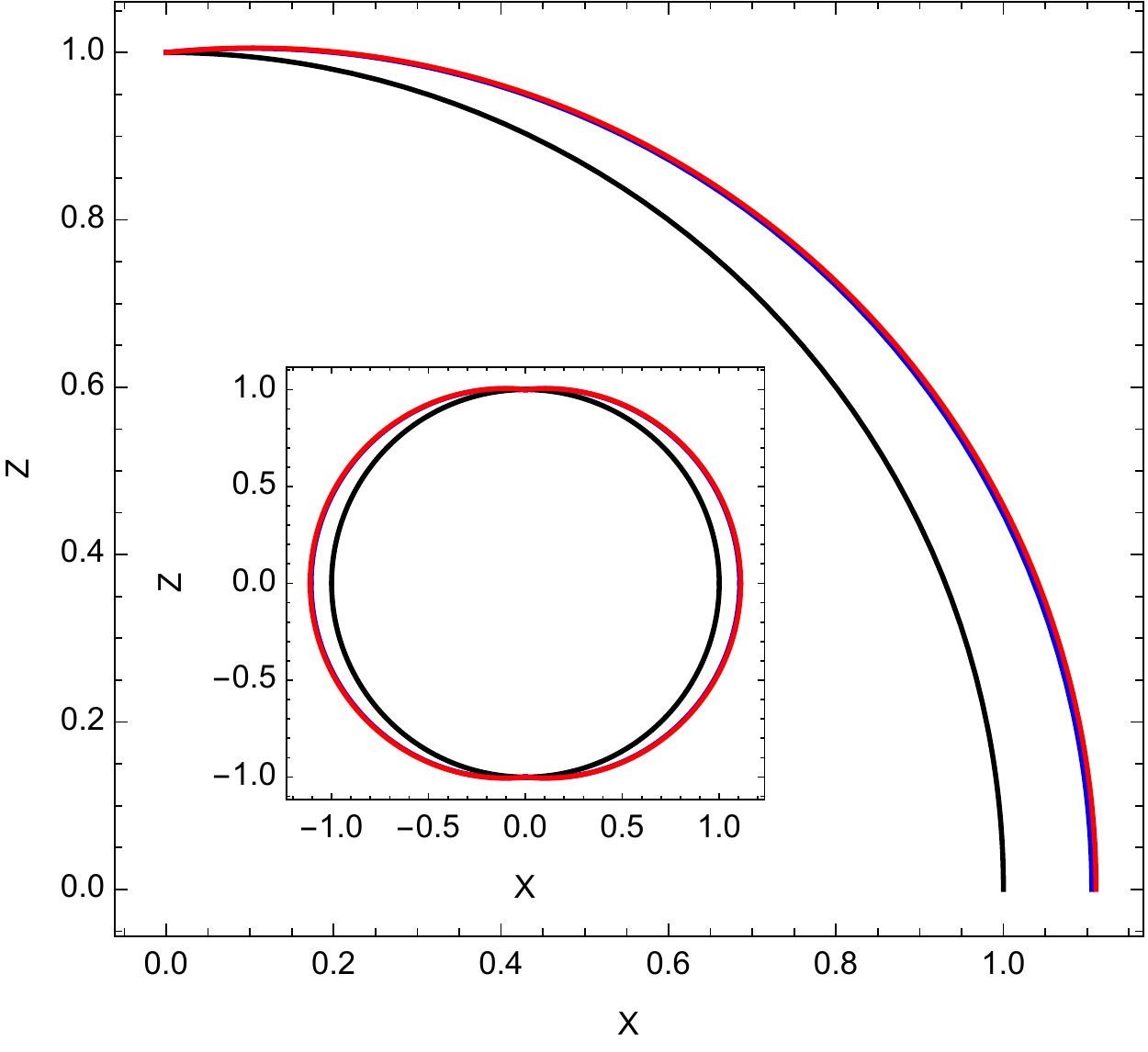}\hfill
  \caption{Ergosphere for a EsGB dilaton black hole with $\gamma=1$, $\chi=0.1$ and $\alpha/M^2=1.15$ (red), together with the ergosphere of a Kerr black hole with the same $\chi$ (blue). The event horizon for both is presented in black.}
  \label{fig:ergo_EsGB}
\end{figure}

\begin{figure}[]
  \centering
      \includegraphics[width=0.7\textwidth]{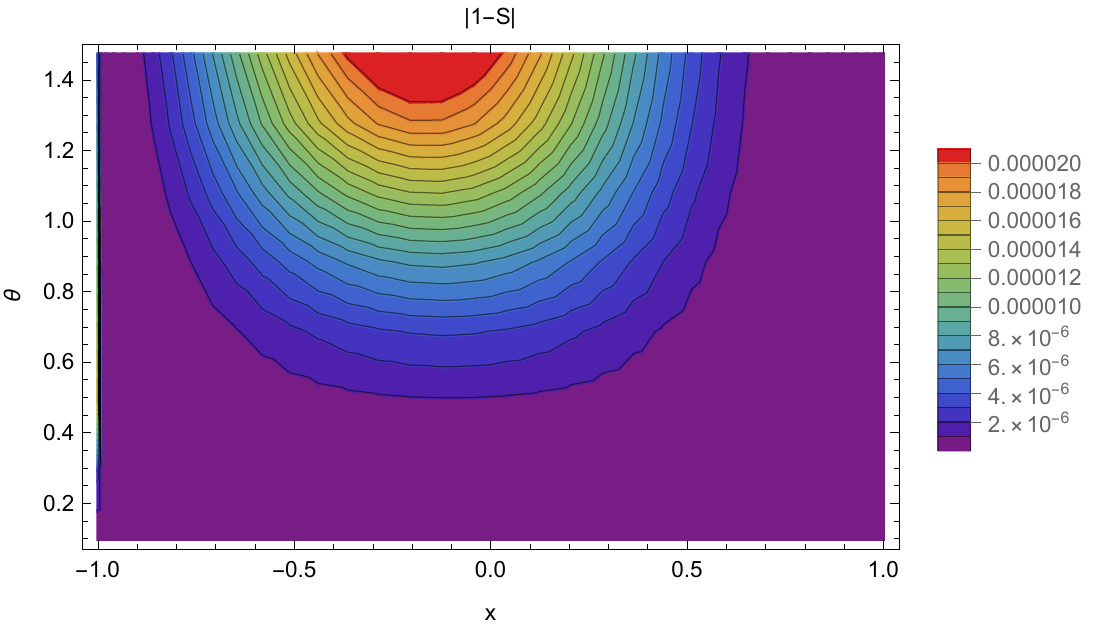}
  \caption{$|1-S|$ plotted as a function of $x$ and $\theta$, where S is the speciality index defined in Eq. \eqref{eq:specindex}, for a EsGB dilaton black hole with $\gamma=1$, $\chi=0.1$ and $\alpha/M^2=1.15$. The non-vanishing value of $|1-S|$ demonstrates that the spacetime is Petrov type I.}
  \label{fig:petrov_EsGB}
\end{figure}

To further demonstrate the capabilities of our code, in the following we present some results for the physical properties of EsGB black holes. A plot of the ergoregion for a dilaton black hole with $\gamma=1$, $\chi=0.1$ and $\alpha/M^2=1.15$ can be found in Fig. \ref{fig:ergo_EsGB}.
In Fig. \ref{fig:petrov_EsGB} we plot $|1-S|$ as a function of $x$ and $\theta$, where S is the speciality index defined in Eq. \eqref{eq:specindex}, for the same EsGB black hole as before, where we can observe that the spacetime is not algebraically special, being Petrov type I. Spinning EsGB black holes were always observed to be Petrov type I.\footnote{With our numerical setup, a Kerr black hole typically yields values of $|1-S|$ on the order of $10^{-15}$ everywhere, in good agreement with the fact that it is Petrov type D.}

\begin{figure}[]
  \centering
      \includegraphics[width=0.5\textwidth]{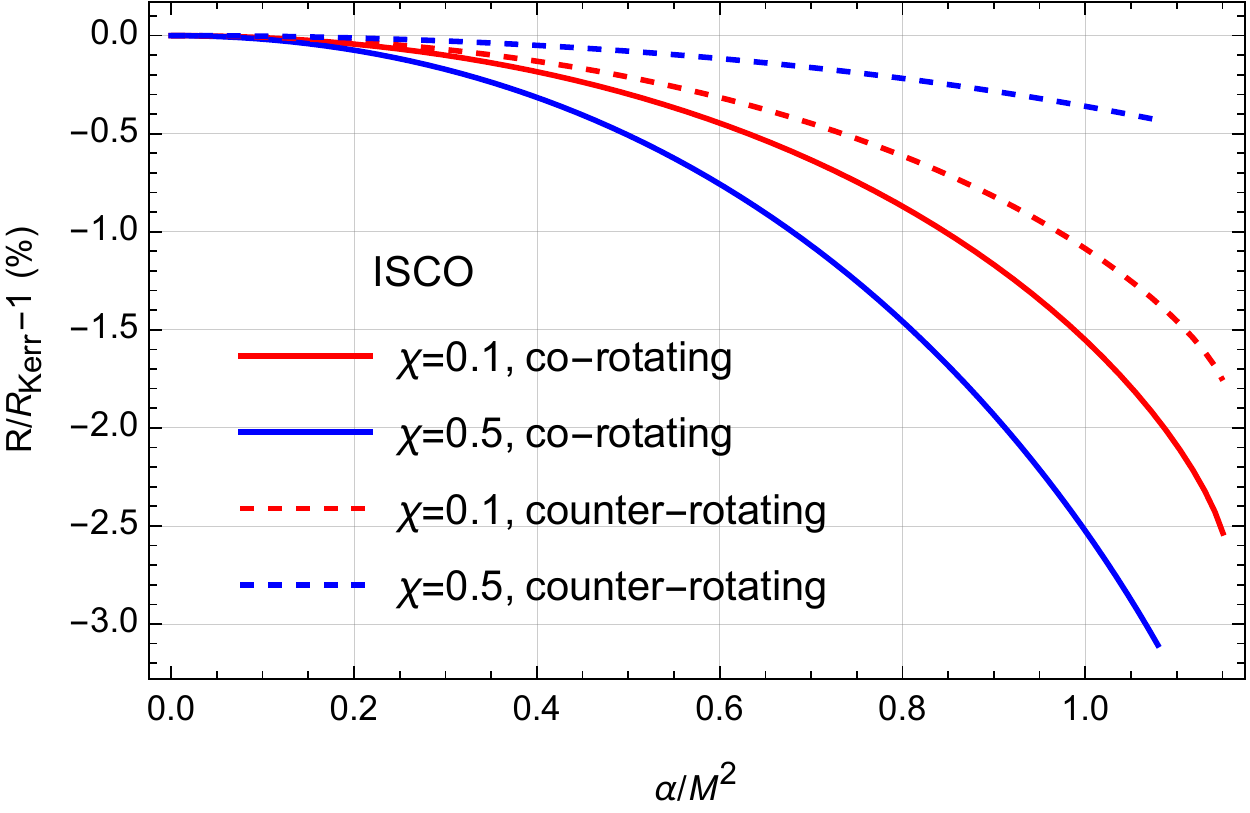}\hfill
      \includegraphics[width=0.5\textwidth]{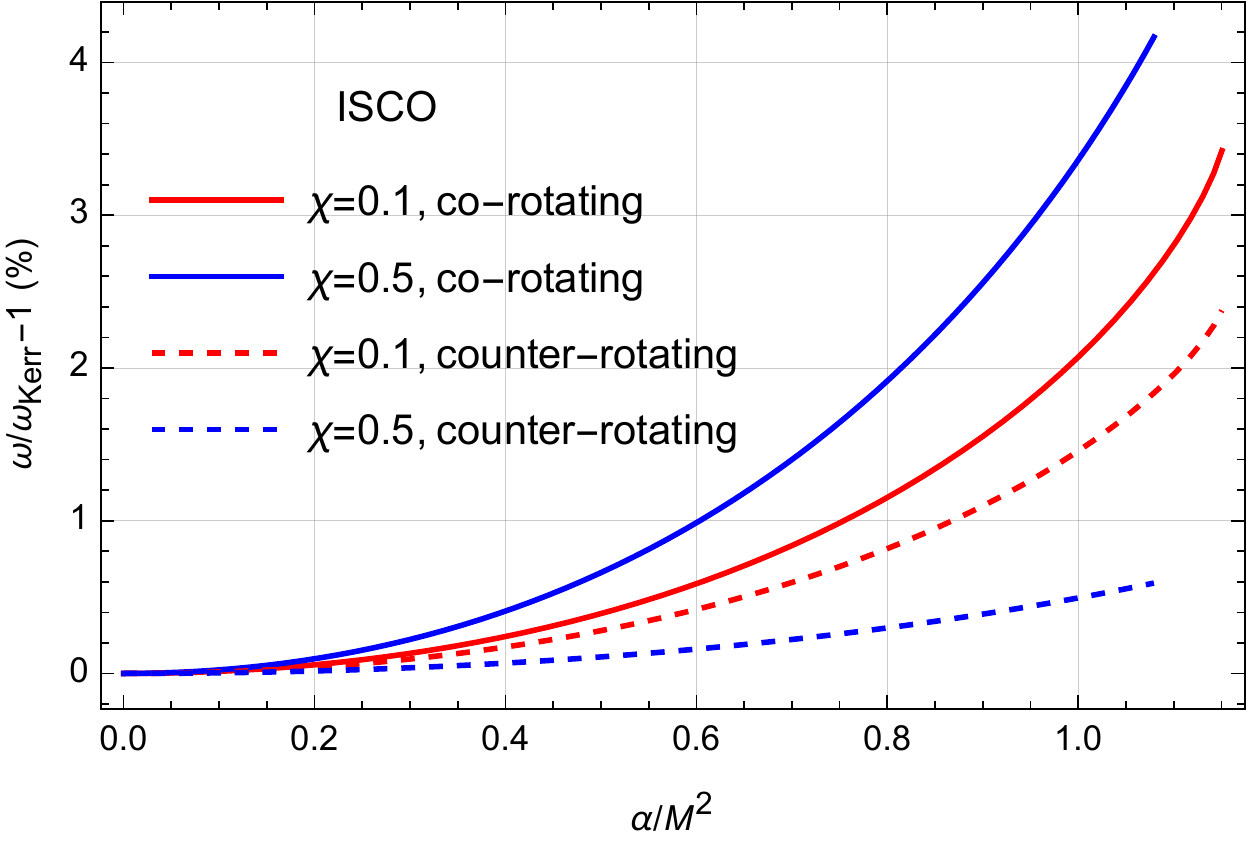}\vfill
      \includegraphics[width=0.5\textwidth]{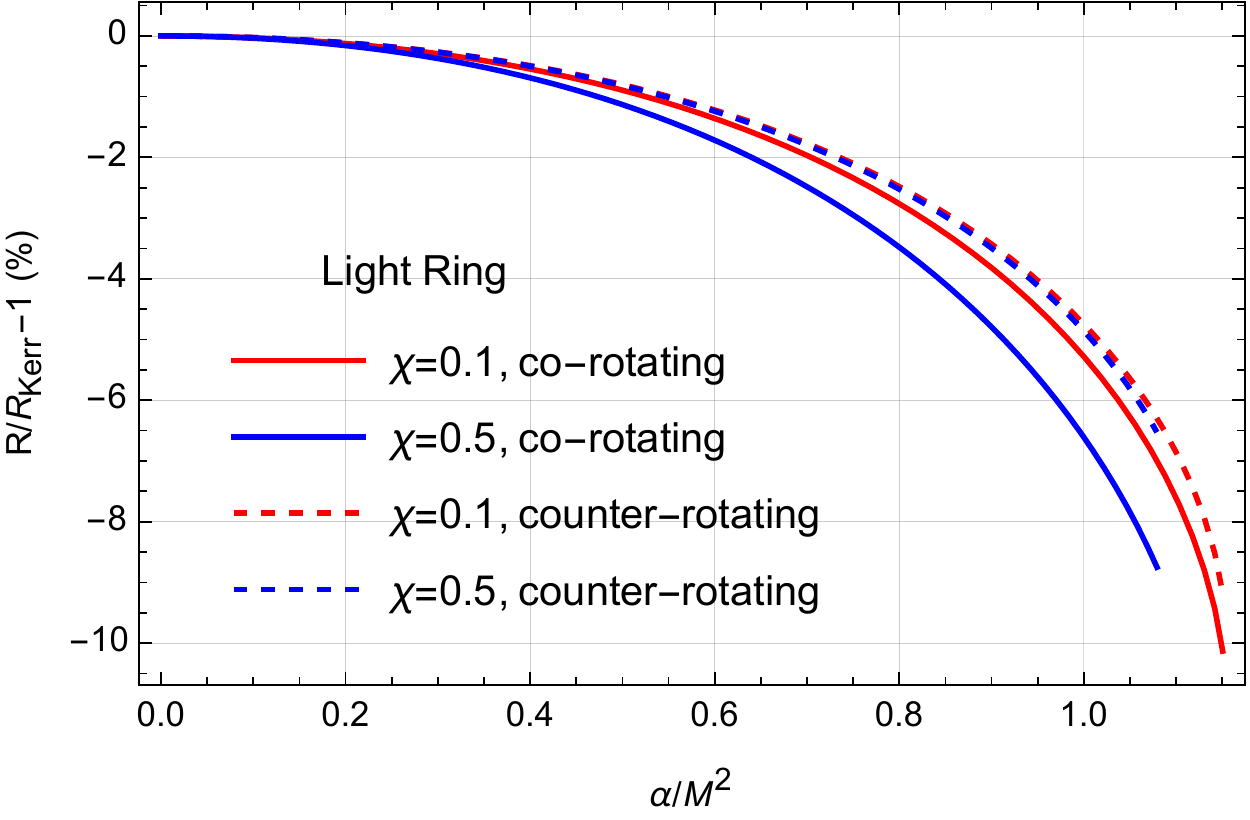}\hfill
      \includegraphics[width=0.5\textwidth]{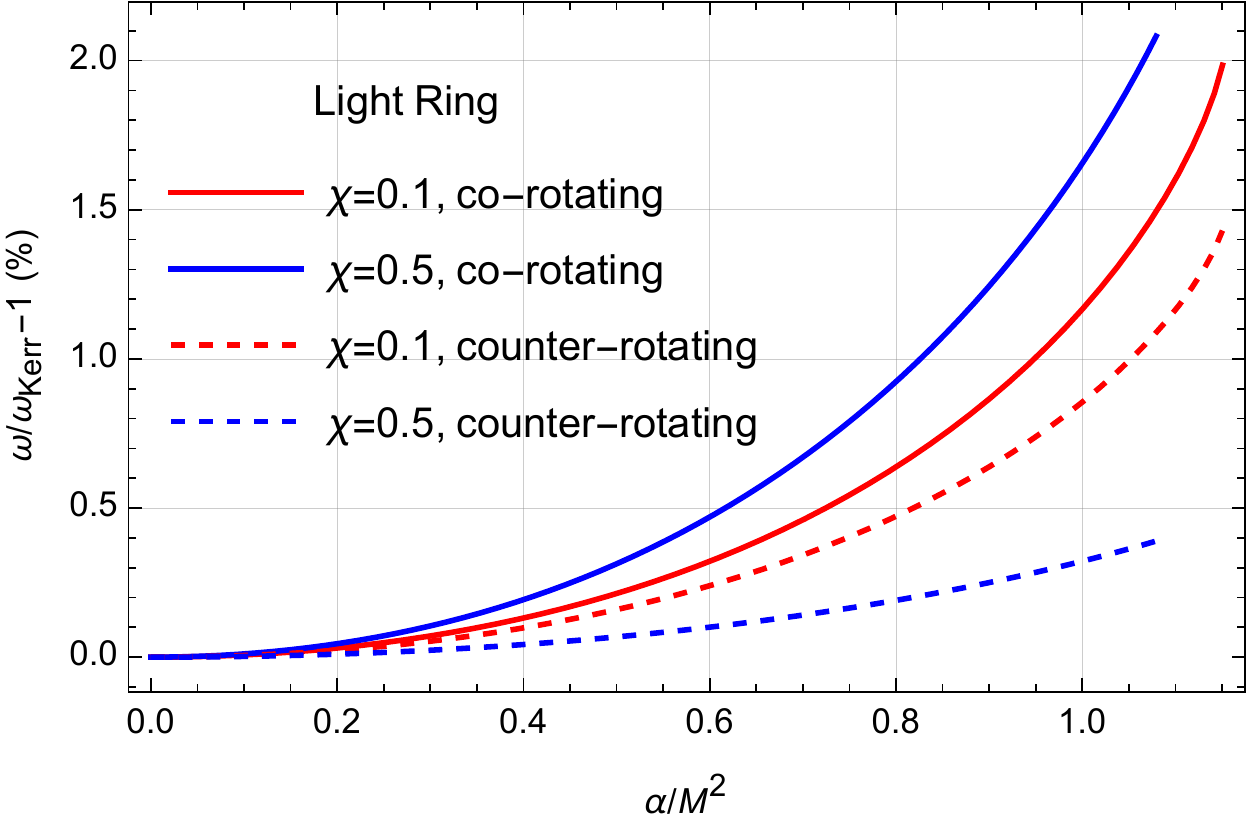}
  \caption{Comparison between EsGB dilaton ($\gamma=1$) and Kerr black holes with the same $\chi$ (and $M$) regarding the perimetral radius and angular frequencies at the ISCO (top) and light ring (bottom) as a function of $\alpha/M^2$, in a part of the domain of existence of solutions.}
  \label{fig:ISCO_EsGB}
\end{figure}

The perimetral location and angular frequencies at the ISCO and light rings of EsGB dilaton black holes ($\gamma=1$) are compared with those of a Kerr black hole (with the same $\chi$ and $M$) in Fig. \ref{fig:ISCO_EsGB}. Note that we have neglected any couplings between the dilaton and matter (see e.g. \cite{Kleihaus:2015aje,Pani:2009wy}). We have compared our results in the static and slowly rotating cases with those in Ref. \cite{Pani:2009wy}, observing remarkable agreement (in the appropriate setup). From Fig. \ref{fig:ISCO_EsGB} we observe differences of a few percent in most cases, with the most drastic differences occurring for the location of the co-rotating light ring due to its proximity to the horizon. The qualitative behaviour is as follows: the perimetral radius of both the ISCO and the light ring decreases with $\alpha/M^2$, and the opposite happens for the angular frequencies\footnote{We note that, similarly to Refs. \cite{Sullivan:2019vyi,Sullivan:2020zpf}, positive coordinate shifts in the location of the ISCO/light ring were observed. These are, however, not physically relevant and the perimetral radius should be used, where negative shifts are observed.}. Co-rotating orbits are most affected, and black hole spin enhances the differences of co-rotating orbits with respect to the Kerr case.

\subsection{Comparison with other codes}
Similar codes to the one we have developed in this chapter are scarce. Indeed, most of the numerical studies of spinning black holes in modified theories of gravity make use of  the non-publicly-available FIDISOL/CADSOL solver \cite{Solver1,Solver2,Solver3}, which implements a finite difference method together with the root finding Newton-Raphson method. The solver is written in Fortran and was first developed in the eighties. Works that use the FIDISOL/CADSOL solver can be found e.g. in Refs. \cite{Herdeiro:2014goa, Herdeiro:2015gia, Herdeiro:2016tmi,Delgado:2020rev, Kleihaus:2015aje,Kleihaus:2011tg,Herdeiro:2020wei,Berti:2020kgk,Cunha:2019dwb}. Some of these works have applied the FIDISOL/CADSOL solver in studies of EsGB gravity, much like we did here. However, they report an error of order $\mathcal{O}\left(10^{-3}\right)$, as estimated by the Smarr relation. In the appendix of Ref. \cite{Delgado:2022pwo}, the author gives a comprehensive overview of the FIDISOL/CADSOL solver, benchmarking it against the Kerr solution, with results again showing errors several orders of magnitude higher than those presented in Fig. \ref{fig:kerr_domain}.

More recently, in Ref.~\cite{Sullivan:2020zpf} the authors developed the \emph{eXtreme Partial Differential Equations Solver} (XPDES) code which is publicly available, to address similar problems. The code is written in C language, and implements a finite difference method to solve the field equations, similarly to the FIDISOL/CADSOL package. It makes use of the software \emph{Maple} to export the field equations to many large C programming files. 
Ref. \cite{Sullivan:2020zpf} does not discuss errors as estimated by Smarr relations, instead, they (also) benchmark their code against the Kerr solution, and compare their EsGB results to perturbative solutions, finding good agreement. They report typical maximum errors on obtaining the Kerr solution of $\mathcal{O}\left(10^{-6}\right)$, which represents a good improvement when compared with the FIDISOL/CADSOL package, especially given that the XPDES code is open-source and publicly available. 

Our code is written in the Julia programming language, which when compared with complied languages such as C code makes it logistically easier to use and adapt, and to implement 
new models. 
In our implementation the field equations and boundary conditions are  written in a very simple way. For example, the boundary condition
\begin{equation*}
  f-2\partial_x f = 0,
\end{equation*}
is written as a residual in code language as
\begin{equation*}
  f - 2*dfdx.
\end{equation*}
The code is memory efficient and fast, making use of pseudospectral methods as explained above, with solutions to the field equations being obtained in the order of a few seconds in laptop-class computers. In our (limited) comparisons with the XPDES code, we found that where our code took only a few seconds the XPDES code would take minutes to achieve a lower accuracy.

The results of this section, for example in Figs. \ref{fig:kerr_domain} and \ref{fig:EsGB_domain_example_exp}, show that the accuracy of our code is many orders of magnitude better than the accuracy presented by either the FIDISOL/CADSOL package or even the XPDES code.

Once a solution to the field equations has been obtained, our code has built-in functions to compute all the physical properties of the black holes discussed in section \ref{sec:physical_properties}, therefore allowing for a simple and comprehensive study of different models. 

\section{Conclusions}

In this paper we have reviewed the spectral method for solving differential equations and subsequently argued that such methods are ideal for finding stationary and axisymmetric black hole solutions in modified theories of gravity. In particular, they allow complicated field equations and boundary conditions to be implemented in a straightforward manner. We  showed how this can be done, and have implemented the method in a new code. To show it in action, and to benchmark its  performance against other codes, we applied the code in the GR setting, and verified that the solution found is extremely close to the known Kerr black hole. We then applied it to a popular set of modified theories of gravity, Einstein-scalar-Gauss-Bonnet gravity, where it is known that black hole solutions different from Kerr exist. In this latter setting we verified the accuracy using analytical expressions that should hold identically. We found that even in the Gauss-Bonnet setting our code took just seconds to find accurate spinning black hole solutions. 

Within the code we have also implemented many built in functions to calculate black hole properties of physical interest. In the future, obtained solutions together with these functions could be used to study a huge range of phenomena observational interest. Other possible studies include the quasi-normal modes of black hole mergers (hence permitting realistic data analysis with Bayesian methods), the electromagnetic emission from accretion disks, black hole shadows, and our code's solutions could also be used as seed solutions for numerical evolutions. Given that the code has been completed only recently, we have, however, not yet applied it widely. Although a first application in research work to EsGB theories is contained in Ref.~\cite{Fernandes:2022kvg}. In the future we hope to apply the code to other theories, such as the so called regularized 4D-Einstein-Gauss-Bonnet gravity theory \cite{Glavan:2019inb,Fernandes:2022zrq,Lu:2020iav,Kobayashi:2020wqy,Fernandes:2020nbq,Hennigar:2020lsl,Fernandes:2021dsb,Aoki:2020lig,Fernandes:2021ysi} where thus far spinning black holes have not been found, and use it to further understand and constrain such theories using the physical properties described.

\section*{Acknowledgements}
P.F. acknowledges support by the Royal Society Grant No. RGF/EA/180022 and is supported by a Research Leadership Award from the Leverhulme Trust. D.J.M. is supported by a Royal Society University Research Fellowship. 

\begin{appendices}
\section{The Kerr-Newman Black Hole}
\label{ap:KN}
The Kerr-Newman solution solves the Einstein-Maxwell field equations
\begin{equation}
  G_{\mu \nu}=2\left(F_{\mu}^{\phantom{\mu} \alpha} F_{\nu \alpha} - \frac{1}{4} g_{\mu \nu} F_{\alpha \beta} F^{\alpha \beta}\right).
\end{equation}
The Einstein-Maxwell field equations can be obtained with the following action principle
\begin{equation}
    \mathcal{S} = \frac{1}{16\pi} \int d^4x \sqrt{-g} \left(R - F_{\mu \nu} F^{\mu \nu}\right),
\end{equation}
where $F_{\mu \nu} = \nabla_\mu A_\nu - \nabla_\nu A_\mu$ is the Maxwell tensor. With the ansatz of Eq. \eqref{eq:metric} the Kerr-Newman black hole solution reads (in terms of $r_H$, $M$ and $Q$)
\begin{equation}
    \begin{aligned}
        &f_{\mathrm{KN}} = \left(1+\frac{r_H}{r}\right)^2 \frac{\mathcal{A}}{\mathcal{B}},\\&
        g_{\mathrm{KN}} = \left(1+\frac{r_H}{r}\right)^2,\\&
        h_{\mathrm{KN}} = \frac{\mathcal{A}^2}{\mathcal{B}},\\&
        W_{\mathrm{KN}} = \frac{r \left(2 M^2-Q^2\right)+2 M \left(r^2+r_H^2\right)}{r_H r^3 \mathcal{B}}\sqrt{M^2-Q^2-4 r_H^2}
    \end{aligned}
  \label{eq:kerr-newman}
\end{equation}
where
\begin{equation}
    \begin{aligned}
    &\mathcal{A} = \frac{r^2 \left(2 M^2-Q^2\right)+2 M r \left(r^2+r_H^2\right)+\left(r^2-r_H^2\right)^2}{r^4}-\frac{\left(M^2-Q^2-4 r_H^2\right)}{r^2}\sin^2\theta ,\\&
    \mathcal{B} = \left(\mathcal{A} + \frac{ \left(M^2-Q^2-4 r_H^2\right) }{r^2} \sin^2\theta\right)^2 - \frac{\left(r^2-r_H^2\right)^2 \left(M^2-Q^2-4 r_H^2\right)}{r^6} \sin^2\theta,
    \end{aligned}
\end{equation}
together with the four-potential
\begin{equation}
    A_\mu dx^\mu = \left(\tilde{A}_t - \frac{W_{KN}}{r}\left(1-\mathcal{N}\right) \tilde{A}_\varphi \sin^2 \theta \right) dt + \tilde{A}_\varphi \sin^2 \theta d\varphi,
\end{equation}
where
\begin{equation}
    \tilde{A}_\varphi = \frac{Q r \left(1 + \frac{M}{r} + \frac{r_H^2}{r^2}\right) \sqrt{M^2-Q^2-4 r_H^2} }{r^2 \left(1 + \frac{M}{r} + \frac{r_H^2}{r^2}\right)^2+\left(M^2-Q^2-4 r_H^2\right)\cos^2\theta},
\end{equation}
and
\begin{equation}
    \tilde{A}_t = \Phi - \frac{Q r \left(1 + \frac{M}{r} + \frac{r_H^2}{r^2}\right)}{r^2 \left(1 + \frac{M}{r} + \frac{r_H^2}{r^2}\right)^2+\left(M^2-Q^2-4 r_H^2\right)\cos^2\theta} + \frac{W_{KN}}{r}\left(1-\mathcal{N}\right) \tilde{A}_\varphi \sin^2 \theta,
\end{equation}
where $Q$ the electric charge and $\Phi$ the electrostatic potential (which can be chosen such that $\tilde{A}_t|_{r_H} = 0$). This particular choice of functions $\tilde{A}_t$ and $\tilde{A}_\varphi$ for the vector potential is such that they are optimised for a numerical setup such as ours.

The total angular momentum (per unit mass), $a$, of the solution is related to $M$, $Q$ and $r_H$ via
\begin{equation}
  r_H = \frac{\sqrt{M^2-a^2-Q^2}}{2} \equiv \frac{M}{2}\sqrt{1-\chi^2-q^2},
\end{equation}
where we have defined the dimensionless charge
\begin{equation}
  q \equiv Q/M.
\end{equation}
The electric charge can be read off the asymptotic decay of the temporal part of the four potential
\begin{equation}
  \tilde{A}_t = \Phi -\frac{Q}{r} + \mathcal{O}\left(r^{-2}\right).
\end{equation}
The Kerr-Newman black hole obeys the well-known Smarr relation
\begin{equation}
    M = 2 T S + 2\Omega_H J + \Phi Q.
\end{equation}
Note that the Kerr-Newman black hole in the quasi-isotropic coordinate system presented in Eq. \eqref{eq:metric} can be obtained from the standard textbook Boyer-Lindquist coordinates solution with the radial coordinate transformation
\begin{equation}
  r_{BL} = r + M + \frac{M^2 - a^2 - Q^2}{4r} = r \left(1 + \frac{M}{r} + \frac{r_H^2}{r^2}\right).
\end{equation}
Details about marginal stable circular orbits in the Kerr-Newman case can be found in Refs. \cite{LRISCO_KN, Wang:2022ouq}.
\end{appendices}

\section*{Bibliography}

\bibliography{biblio}

\end{document}